\documentstyle[epsfig,prd,preprint,aps,eqsecnum]{revtex}

\begin{document}
\tighten
\reversemarginpar

\title{Quantum Mechanics of a Black Hole}
\author{Gilad Gour\thanks{E-mail:~gour@cc.huji.ac.il}} 
\address{Racah Institute of Physics, Hebrew University, 
Givat Ram, Jerusalem~91904, ISRAEL.}

\maketitle

\begin{abstract}
Beginning with Bekenstein, many authors have considered a uniformly
spaced discrete quantum spectrum for black hole horizon area. It is also
believed that the huge degeneracy of these area levels corresponds to the
notion of black hole entropy. Starting from these two assumptions we here
infer the  {\it algebra} of a Schwarzschild black hole's observables. This
algebra then serves as motivation for introducing in the system's Hamiltonian
an interaction term.  The interaction contains the horizon area operator,
which is a  {\it number} operator, and its canonical conjugate, the {\it
phase} operator. The Hawking radiation from a Schwarzschild black hole is seen
to be a consequence of an area--phase interaction. Using this interaction  we
have reproduced the semi--classical result for the Hawking radiation power. 
Furthermore, we show that the initial state of the black hole determines the
nature of its development. Thus, a state which is an area eigenstate describes
a static eternal black hole, but a coherent state describes a radiating black
hole. Hence, it is the observer's initial knowledge or uncertainty about the
horizon area which determines the evolution. 
\end{abstract}



\section{Introduction}

The first quarter of this century yielded  two revolutionary theories, quantum
mechanics and general relativity; they have changed drastically the way we
think about the physical world. Since then many efforts have been made to
unify these theories, and today we are presented with a variety of
generic ``quantum gravity'' theories.  Let us mention canonical quantum
gravity~\cite{Kuchar}, perturbative quantum gravity~\cite{Weinberg},
superstring and D-brane theory~\cite{dbrane}, loop (nonperturbative) quantum
gravity~\cite{AshKra} and others. All these are intricate theories: even
comparison of their predictions is a difficult task.  In view of the dearth of
concrete experimental results bearing on the quantum aspects of gravity, one
wonders whether it would not be more to the point to construct a theory which
emphasizes some particular gravitational system where circumstances allow a
guess as to the general nature of quantum effects.  Black holes suggest
themselves as a good place to start: they involve strong gravitational fields,
have properties reminiscent of localized particles, and results from
quantum field theory [principally Hawking's radiance (HR)] suggest that
quantum effects are central. Thus an approach to black hole
quantization has emerged~\cite{Bek74,Mu86,BekMu95,BekLectures} which starts
off from hints garnered in classical black hole physics, and proceeds to
quantum ideas.

This approach hails back to the realization by Christodoulou~\cite{Chri}, 
Hawking~\cite{Haw71} and Penrose and Floyd~\cite{PenFly} that transformations 
of a black hole (BH) generically have an irreversible character. One of its
conclusions is that the horizon area tends to grow, a rule which served as
motivation for Bekenstein~\cite{Bek73} to associate entropy with horizon area
$A$  according to (we insert the modern value of the coefficient) 
\begin{equation}
S=A/(4{\cal L}^{2}_{p})
\label{int}
\end{equation} 
(${\cal L}_{P}$ and ${\cal M}_P$ will denote the Planck length and mass,
respectively). Christodoulou also showed that there are some special BH
reversible processes which leave BH area unchanged.  This has led
Bekenstein~\cite{Bek74} to remark that the  horizon area of a non--extremal
BH behaves as a classical {\it adiabatic invariant} (for support of this conjecture see
also Bekenstein~\cite{Bek98} and Mayo~\cite{Mayo}). According to Ehrenfest's
principle~\cite{Eren} any classical adiabatic invariant corresponds to a
quantum entity with discrete spectrum. Thus Bekenstein conjectured that the
spectrum of the horizon area of a (non--extremal) BH should be quantized.
Today the idea of a discrete eigenvalue spectrum for the horizon area is also
supported by the work of Ashtekhar and others~\cite{AshKra}.

In support of a conjectured uniform spacing of the area eigenvalues,
Bekenstein noted~\cite{Bek74,BekLectures} that the assimilation of a neutral
test particle into a BH increases the horizon area by at least $\upsilon {\cal
L}_{P}^{2}$, where $\upsilon$ is a dimensionless constant of order unity.  Also
Hod~\cite{Hod1} have shown independently that the same kind of lower bound on
the horizon area increase applies to the BH assimilation of a {\it charged}
particle.  The fact that the lower bound turns out to be independent of the BH
parameters (mass, charge and  angular momentum) is in harmony with a uniformly
spaced area spectrum:
\begin{equation}
a_{n}=\upsilon {\cal L}_{P}^{2}(n+\eta);\;\;\;0\leq\eta <1;\;\;\;n=0,1,2,...
\label{intro}
\end{equation}
Here $a_{n}$ is the area eigenvalue. In the present work we shall set $\eta
=0$ for simplicity. 

The spectrum in Eq.~(\ref{intro}) together with the classical relation 
$A=16\pi M^{2}$ (we use gravitational units in which $G=c=1$) implies that 
for $n\gg 1$ the energy (mass) levels $M_{n}$ of an isolated Schwarzschild BH
have the form $M_{n}=m_{0}\sqrt{n}$; here $m_{0}=\sqrt{\upsilon\hbar/16\pi}$.
This kind of spectrum is compatible with Bohr's correspondence principle. That
is, for large $n$ values the 
$n\rightarrow n-1$ transition frequency
\begin{equation}
\omega_{0}\equiv\frac{M_{n}-M_{n-1}}{\hbar}=\frac{\upsilon}{32\pi M}
\end{equation}
agrees with the classical BH oscillation frequencies which scale as $M^{-1}$.

The energy $\hbar\omega_{0}$ emitted in the transition $n\rightarrow n-1$ is
reminiscent of the Hawking temperature. Hawking~\cite{Hawk75} has shown that
a steady emission of particles to infinity originating in pair creation in
the gravitational field follows when a BH is formed by gravitational
collapse. He associated with the BH a temperature $T_{BH}=\hbar/8\pi M$ 
[compatible with the entropy given in Eq.~(\ref{int})] which characterizes
the thermal spectrum of  the emitted particles. Note that $T_{BH}$
and $\hbar\omega_{0}$ are of the same order. Hence, the idea that the
temperature  represent the average energy of the emitted particles is
compatible with a mass spectrum $M\sim\sqrt{n}$.  
 
Following Bekenstein's conjecture about the mass spectrum,  a number of
authors (see full references in~\cite{Kast97} and \cite{BekLectures}) have
obtained  the selfsame spectrum from a wide variety of arguments.  For
example,  Vaz and Witten~\cite{CenLo98} examined the Wheeler-DeWitt equation
for a static, eternal Schwarzschild  BH in Kucha\v{r}-Brown variables, and
obtained the mass spectrum $M_{n}\propto\sqrt{n}$. Louko and
M\"akel\"a~\cite{JorJar96} have proposed a Hamiltonian  operator for the
Schwarzschild BH, and obtained the same spectrum.  Kastrup~\cite{Kast96}
explained how such a spectrum may be understood in the framework of canonical
quantization of a purely gravitational spherically symmetric
system. Carlip and Teitelboim~\cite{CarTe95} used the standard
(Euclidean) action principle for the gravitational field to show that the
horizon area and the  opening angle at the horizon are canonical
conjugates. The opening angle is bounded (in the Euclidean sector). This is
compatible with the idea that the area operator is represented by a number
operator (see Eq.~(\ref{intro}), since its canonical conjugate is a {\it
phase} operator~\cite{Lyn95}.  This similarity is further evidence in favor of
an equally  spaced area spectrum. 

The energy levels $M_{n}$ are expected to be degenerate. Denoting the
degeneracy by $g(n)$ and identifying $\ln g(n)$ with the entropy given in
Eq.~(\ref{int}), Bekenstein and Mukhanov~\cite{BekMu95} found that
$\upsilon=4\ln k$  [$\upsilon$ is the dimensionless constant in 
Eq.~(\ref{intro})],
or equivalently $g(n)=k^{n}$ where $k=2,3,4,\cdots$ They chose $k=2$ for
simplicity; recently Hod~\cite{Hod2}  has given evidence in favor of
the choice $k=3$.

By contrast to the approach just discussed, quantum loop gravity determines a
non--uniformly spaced area spectrum. Ashtekhar and Krasnov claim~\cite{AshKra}
that the area spacing approaches zero exponentially for large quantum numbers.
Hence, the area spectrum they propose is directly compatible with the
semi--classical prediction by Hawking~\cite{Hawk75} that the emitted spectrum
is smooth. The present paper shows that other features of Hawking radiation
emerge naturally from the uniformly spaced area spectrum hypothesis. For
example, we show that the rate of change of the mass of a BH (not near the
Planck scale)  agrees with that of the semi-classical  calculation by
Page~\cite{Pag76} up to the second order in time. Thus, although the uniformly
spaced area spectrum is inimical to the smooth Hawking spectrum, it is not at
variance with equally important features of a semi--classical BH.

It is clear that the Hawking temperature is a quantum effect ($T_{BH}$ is
proportional to Planck's constant $\hbar$), and a deep understanding of it 
requires a quantum theory of BHs.   In this paper a quantum model for
Schwarzschild BHs is proposed.  We start from  two axioms: (a) that the
horizon area spectrum is equally spaced, and (b)  that the degeneracy of level
$n$ is $k^{n}$.  Surprisingly, these two axioms enable us to obtain the
algebra of the Schwarzschild BH observables, and even the form of the  Hamiltonian. The
algebra obtained is the algebra of angular momentum with the restriction that
the absolute value of the ``angular momentum", here called the {\it
hyperspin}, can have only the values $(k^{n}-1)/2$. This last limitation
motivates us  to introduce an interaction term which allows transitions 
only between allowed values of the hyperspin.
 
The $z$--projection of the hyperspin represents the internal degree of freedom 
of the BH [this degree of freedom corresponds to the entropy given
in Eq.~(\ref{int})].  According to the ``no hair" theorems~\cite{NoHair}, an
observer located at large  distance from the BH is not able to distinguish
between different states of the internal  degrees of freedom. Hence, we find it 
necessary to introduce a horizon projection operator which filters out the
hyperspin  degree of freedom, but leaves that corresponding to the mass of the
BH. This corresponds to the conception that most of the information about the
BH interior is hidden (by the horizon) from  an exterior observer.

The Hilbert space which describes the states that the observer at infinity
can measure is spanned by area (mass) eigenstates. We show in this paper that
the  initial state determines the nature of the dynamics of the BH.  For
example, if the initial state is chosen to be an area eigenstate, it
describes a static eternal BH. On the contrary, if the initial state is
chosen to be a coherent state,\footnote{Coherent states are the ones that
most closely imitate the classical results for a simple harmonic oscillator.}
it reproduces the semi-classical result~\cite{Pag76,BekMu95} for the Hawking
power. That is, it is our initial knowledge (uncertainty) about the horizon
area which determines the nature of the dynamics. A BH formed by
gravitational collapse generally has uncertain horizon area, and thus it
looses mass, whereas an eternal BH, with a definite mass and area, is static.

This paper is organized as follows. In section~\ref{sec:ta} we construct the
hyperspin algebra and characterize the observable Hilbert space. In
section~\ref{sec:hamin} we define the mass operator, and then find the general
form of the Hamiltonian of a Schwarzschild BH. We also review the quantum
phase problem in general, and then define the  phase of a BH. In
section~\ref{sec:tis} we discuss the initial states to be  considered, and in
section~\ref{sec:hr} we use them to compare semiclassical properties of the HR
with our  model's predictions. Finally, in section~\ref{sec:summary}, we
present our summary and conclusions.  

\section{Black Hole Algebra}\label{sec:algebra}

The classical Schwarzschild BH is characterized by only one parameter,
its mass or equivalently its horizon area. Since the horizon area
is an adiabatic invariant and is suggested to be quantized (as discussed in
the Introduction), we shall focus on it.
Hence, the first step should be to investigate the 
horizon area operator. In this work we shall not construct 
the area operator from more fundamental quantities~\cite{AshKra}, even though 
this might illuminate our perception of the internal structure of the BH.
Instead, we shall use the properties of the area operator which follow
from general arguments (see the Introduction). We now turn to construct the
algebra.

\subsection{The Algebra}\label{sec:ta} 

Let ${\bf A}$ represent the horizon area operator
(boldface is used everywhere to denote operators). Motivated by the arguments
discussed in the Introduction we shall start from the following two 
axioms:\\
(1) The eigenvalues of ${\bf A}$ are $a_{n}=a_{0}n$
where $n=0,1,2...$,   $a_{0}=4{\rm ln}k{\cal L}_{P}^{2}$ is a positive
universal constant, and $k$ is an
integer greater then one.\\ (2) There are exactly $k^{n}$ independent
eigenstates of ${\bf A}$  with eigenvalue $a_{n}$, i.e. the degeneracy of a
state with area $a_{n}$ is $k^{n}$.

From the second axiom it follows that there is at least 
one more operator which commutes with ${\bf A}$. Let ${\bf G}$ represent 
the set of {\it all} independent 
operators which commute with ${\bf A}$. Typically, 
\begin{equation}
[{\bf A},{\bf G}]=0,
\label{bcr}
\end{equation}
and the set $\{ {\bf A},{\bf G} \}$ of ``observables" is maximal in the 
sense that if ${\bf S}$ is another observable which commutes with both ${\bf
A}$ and ${\bf G}$, then ${\bf S}=f({\bf A},{\bf G})$ where $f$ is a function
of ${\bf A}$ and ${\bf G}$. We shall call ${\bf G}$ a ``secret" observable
since it describes the internal degrees of freedom of the BH.

From the second axiom or from Eq.~(\ref{bcr}) it follows that we can
simultaneously diagonalize ${\bf A}$ and ${\bf G}$. Hence we shall
denote the states as $|n,m\rangle$ such that
\begin{eqnarray}
{\bf N}|n,m\rangle & = & n|n,m\rangle\;\;\;\;\;\;n=0,1,2...\nonumber\\
{\bf G}|n,m\rangle & = & m|n,m\rangle\;\;\;\;\;\;m=1,2,...,k^{n}
\label{bev}
\end{eqnarray}
where ${\bf N}\equiv a_{0}^{-1}{\bf A}$ is a dimensionless number 
operator.
The eigenvalues $m$ of ${\bf G}$ are limited by $n$ since we want a 
degeneracy of $k^{n}$. We have assumed that $m$ is an integer; if not,
instead of ${\bf G}$ we shall choose the ``secret'' operator to be 
$f({\bf G})$, where $f(m)=1,2,...,k^{n}$ for all the different $k^{n}$ values
of $m$.

Now, since ${\bf G}$ has a discrete spectrum $m=1,2,...,k^{n}$,
we shall introduce the {\it annihilation operator} ${\bf g}$ and the 
{\it creation operator} ${\bf g^{\dag}}$ corresponding to ${\bf G}$:
\begin{equation}
[{\bf G},{\bf g}]=-{\bf g}\;\;\;\;\;\;[{\bf G},{\bf g^{\dag}}]={\bf g^{\dag}}.
\label{Gcr}
\end{equation}
As we shall see shortly, under certain conditions ${\bf g}$ 
and ${\bf g^{\dag}}$ have to satisfy a certain commutation relation. 
In order to single out an algebra, we have to demand that 
$[{\bf g},{\bf g^{\dag}}]$ be linear in ${\bf G}$. In order to derive this
commutation rule we shall first guess some of the properties of ${\bf g}$ and
${\bf g^{\dag}}$ from physical arguments. 
First we do not want ${\bf g}$ and ${\bf g^{\dag}}$ to change $n$ when they 
act on states with a definite area eigenvalue $n$. This means that transitions
among the internal degrees of freedom ($m\rightarrow m'$) do not have to
change the area of the BH. Thus, the area operator ${\bf N}$ commutes with
${\bf g}$ and ${\bf g^{\dag}}$.

Next, from Eq.~(\ref{Gcr}) we have by analogy with the harmonic oscillator
\begin{equation}
{\bf g^{\dag}}|n,m\rangle\propto |n,m+1\rangle.
\label{gdag}
\end{equation}
Therefore ${\bf g^{\dag}}|n,k^{n}\rangle=0$ and thus also 
${\bf gg^{\dag}}|n,k^{n}\rangle=0$.
It is clear from Eq.~(\ref{Gcr}) that 
$[{\bf G},{\bf gg^{\dag}}]=0$ and also that
$[{\bf N},{\bf gg^{\dag}}]=0$. But the set 
$\left( {\bf N},{\bf G} \right)$
is maximal and thus, since ${\bf gg^{\dag}}$ is an observable,  
${\bf gg^{\dag}}=h_{1}({\bf N},{\bf G})$ where $h_{1}$ is a  
function of ${\bf A}$
and ${\bf G}$. Hence, from the fact that 
${\bf gg^{\dag}}|n,k^{n}\rangle=0$ we find that 
$h_{1}(n,k^{n})=0$. Similarly we find that
\begin{equation}
{\bf g^{\dag}g}|n,1\rangle=0,
\label{ab}
\end{equation}
where ${\bf g^{\dag}g}=h_{2}({\bf N},{\bf G})$. It follows from Eq.~(\ref{ab}) that
$h_{2}(n,1)=0$. Collecting all of these facts we find that ${\bf g}{\bf
g}^{\dag}$ and ${\bf g}^{\dag}{\bf g}$ can be written uniquely  as 
\begin{eqnarray}
{\bf gg^{\dag}} & = & u({\bf N}){\bf G}(k^{\bf N}-{\bf G})\nonumber\\
{\bf g^{\dag}g} & = & u({\bf N})(k^{\bf N}-{\bf G}+1)({\bf G}-1),
\label{ggd}
\end{eqnarray}
where $u({\bf N})$ is some function of ${\bf N}$. In Appendix A we show that 
$u({\bf N})=1$. Thus, we settle on the commutation relation
\begin{equation}
[{\bf g},{\bf g^{\dag}}]=k^{\bf N}-2{\bf G}+1
\label{mcr}
\end{equation}
which is linear in ${\bf G}$. Eq.~(\ref{mcr}) and Eq.~(\ref{Gcr}) seem to 
imply that the set of operators $\{ {\bf G},{\bf g},{\bf g^{\dag}}\}$
plus the identity element span a 4-dimensional Lie algebra which 
constitutes a central extension of the rotation group. The 
representations are supposed to be such that the maximal value of the third
component is just $k^{n}$. The type of the central extension is exhibited
in Eq.~(\ref{mcr}) where the term $(k^{n}+1)/2$ is the central charge 
(a multiple of the identity). However, as we shall see later, we can eliminate
the central charge from Eq.~(\ref{mcr}) by a redefinition of the third 
generator ${\bf G}\rightarrow {\bf G}-(k^{n}+1)/2$.
Furthermore, we shall make it clear under what 
circumstances the commutation rule~(\ref{mcr})
is consistent with the required spectrum of ${\bf N}$ and ${\bf G}$. 
   
\subsubsection{The Matrix Elements}

We turn now to calculate the matrix elements of the 
operators in the algebra. We chose the orthonormal set
of states $\{|n,m\rangle\}_{m=1,2,...,k^{n}}^{n=0,1,2,...}$
as a basis. Thus we have 
\begin{eqnarray}
\langle n'm'|{\bf N}|n,m\rangle & = & n\delta_{n,n'}\delta_{m,m'}\\
\langle n'm'|{\bf G}|n,m\rangle & = & m\delta_{n,n'}\delta_{m,m'}.
\end{eqnarray}
Now, from~(\ref{gdag}) it follows that 
${\bf g^{\dag}}|n,m\rangle=C_{nm}^{+}|n,m+1\rangle$
where
\begin{equation}
|C_{nm}^{+}|^{2}=\langle nm|{\bf gg^{\dag}}|n,m\rangle=
\langle nm|{\bf G}(k^{\bf N}-{\bf G})|n,m\rangle=
m(k^{n}-m).
\end{equation}
In a similar fashion ${\bf g}|n,m\rangle=C_{nm}^{-}|n,m-1\rangle$
were
\begin{eqnarray}
|C_{nm}^{-}|^{2}=\langle nm|{\bf g^{\dag}g}|n,m\rangle & = &
\langle nm|(k^{\bf N}-{\bf G}+1)({\bf G}-1)|n,m\rangle\nonumber\\
& = & (k^{n}-m+1)(m-1)
\end{eqnarray}
and thus we finally obtain for the matrix elements of ${\bf g}$ and 
${\bf g^{\dag}}$:
\begin{eqnarray}
\langle n'm'|{\bf g^{\dag}}|n,m\rangle & = & 
\left(m(k^{n}-m)\right)^{1/2}\delta_{n,n'}\delta_{m',m+1}\nonumber\\
\langle n'm'|{\bf g}|n,m\rangle & = & 
\left((k^{n}-m+1)(m-1)\right)^{1/2}\delta_{n,n'}\delta_{m',m-1},
\end{eqnarray}
where for simplicity we have chosen $C_{nm}^{+}$ and $C_{nm}^{-}$ 
to be a positive real numbers. These matrix elements are reminiscent of the
matrix elements of $J_{-}\equiv J_{x}-iJ_{y}$ and $J_{-}\equiv J_{x}+iJ_{y}$
in the algebra of angular momentum. As we shall see shortly, this is not
accidental.

\subsubsection{The Hyperspin of a Black Hole} 

Using Eq.~(\ref{ggd}) we can find a relation between ${\bf N},{\bf g}$ 
and ${\bf g^{\dag}}$
\begin{equation}
k^{2{\bf N}}=[{\bf g},{\bf g^{\dag}}]^{2}+2\{{\bf g},{\bf g^{\dag}}\}+1
\label{rag}
\end{equation}
where the symbol $\{,\}$ indicates an anticommutator. It is not clear 
{\it a priori}
that building ${\bf N}$ from ${\bf g}$ and ${\bf g^{\dag}}$ in this way is 
consistent
with the required spectrum of ${\bf N}$. In order to clarify this 
point let us
first define three observables:
\begin{eqnarray}
{\bf S_{1}} & \equiv & \frac{1}{2}({\bf g}+{\bf g^{\dag}})
\nonumber\\
{\bf S_{2}} & \equiv & \frac{i}{2}({\bf g}-{\bf g^{\dag}})
\nonumber\\
{\bf S_{3}} & \equiv & {\bf G}-\frac{1}{2}k^{{\bf N}}-\frac{1}{2}.
\label{jjj}
\end{eqnarray}
Using the commutation relations between ${\bf g}$,
${\bf g^{\dag}}$, ${\bf N}$ and
${\bf G}$ we notice that the operators defined in~(\ref{jjj}) satisfy the
standard commutation rules of an angular momentum, that is
\begin{equation}
[{\bf S}_{i},{\bf S}_{j}]=i\varepsilon_{ijk}{\bf S}_{k},
\end{equation}
where $\varepsilon_{ijk}$ is the standard antisymmetric tensor. This angular 
momentum 
algebra is not surprising for an algebra with 
three generators. As we shall see below, the requirement that ${\bf N}$ is a 
number
operator complicates the algebra. 

Let us now rewrite Eq.~(\ref{rag}) in terms of the angular 
momentum operators:
\begin{equation}
k^{2{\bf N}}=4{\bf S}^{2}+1
\label{raj}
\end{equation}
where ${\bf S}^{2}={\bf S}_{1}^{2}+{\bf S}_{2}^{2}+{\bf S}_{3}^{2}=s(s+1)$ 
($s$ is a half integer).
It is clear from Eq.~(\ref{raj}) and the definition of ${\bf S_{3}}$, 
that a simultaneous eigenstate of ${\bf S}^{2}$ and ${\bf S_{3}}$
,which is denoted by $|s,m_{3}\rangle$, is also a simultaneous 
eigenstate of ${\bf N}$ and ${\bf G}$. Explicitly, 
$|s,m_{3}\rangle=|n,m\rangle$ for
\begin{equation}
m=s+m_{3}+1 \;\;\;{\rm and}\;\;\; k^{n}=2s+1.
\label{nms}
\end{equation}

Two features follow from this relation. Firstly, the modulus of the
angular momentum (the ``hyperspin'' of the BH)
is a true observable, whereas its $z$-projection
represents the hidden degrees of freedom of the BH. Secondly,
we can see that $n$ may be an integer, but it may also be 
log$_{k}(2s+1)$ where $s=0,\frac{1}{2},1,\frac{3}{2},2,...$.
Thus, the Hilbert space ${\cal H}$ spanned by the set  
$\{|n,m\rangle\}$ for an integer $n$ is a subspace of the 
``full'' Hilbert space ${\cal H}_{\rm {f}}$ spanned by the set 
$\{|s,m_{3}\rangle\}$. Since we have obtained the angular momentum operators,
one can build the algebra in the opposite direction (starting with the three
angular momentum operators) and ask, what is the physical reason 
that a BH state is restricted to a proper subspace of 
${\cal H}_{\rm {f}}$?
Why can it not have general values of $s$, but only those of
the form $(k^{n}-1)/2$?

These questions motivate us to enter into the dynamics, and
define the Hamiltonian. It is physically clear that transitions between 
states with spin of the form $(k^{n}-1)/2$ and states which
are not of the same form are forbidden. Before starting to construct 
the Hamiltonian, let us introduce the Hilbert space of an 
observer who is located very far a way from the BH.

\subsection{The Observable Hilbert Space}\label{subsec:ohs}

An observer far away from a Schwarzschild BH
is able to measure only the mass of the BH. ``No hair" theorems~\cite{NoHair}
suggest that this is the sole parameter characterizing the BH.
On the other hand, the states in Eq.~(\ref{bev}) are characterized by the two
parameters $n$ and $m$. This raises the question, how can we harmonize between
our states and ``no hair" theorems? In other words, how to include the
Israel-Carter conjecture that ``black holes have no hair" in the Hilbert
space which is spanned by the {\it degenerate} area eigenstates?

At first sight, a description of BHs by means of mixed states with entropy
$S=A/4{\cal L}^{2}_{p}$ instead of pure states (zero entropy) seems to give
the answer  to the question above. Denoting the density matrix by ${\bf
\rho}_{n}$ for each area eigenvalue $n$, we find that
\begin{equation}
{\bf \rho}_{n}\equiv\frac{1}{k^{n}}\sum_{m=1}^{k^{n}}|n,m\rangle\langle n,m|
\label{mix}
\end{equation}
and thus the entropy 
$S\equiv -{\rm Tr}{\bf \rho}_{n}\ln {\bf \rho}_{n}=A/4{\cal L}^{2}_{p}$. This
agrees with Hawking's {\it principle of ignorance}~\cite{Hawki75}: ``the
surface  emits with equal probability all configurations of particles
compatible with the  observers limited knowledge". However, the problem with 
Eq.~(\ref{mix}) is that it suggests that the observer is  able to identify
$|n,m_{1}\rangle$ as a  different state from $|n,m_{2}\rangle$ where
$m_{1}\ne m_{2}$. Thus, Eq.~(\ref{mix}) is not compatible with ``no hair"
theorems. But without a density matrix, how to attribute a BH 
an entropy $S=A/4{\cal L}^{2}_{p}$? In order to resolve this issue we
introduce the concept of a {\it secret} observable. 

Motivated by ``no hair" theorems we claim that the {\it observer} 
is not able to distinguish
between {\it different} states with the same area eigenvalue. Hence there is a
subspace of ${\cal H}$ which represents the observable Hilbert space. In this
subspace for each area eigenvalue there is only one area eigenstate, i.e. 
the area operator is not degenerate in the observable subspace. Now, each
area eigenstate in the observable subspace is a linear superposition of the states 
$\{|n,m\rangle\}_{m=1,2,...,k^{n}}$, that is
\begin{equation}
|n\rangle_{ob}\equiv\sum_{m=1}^{k^{n}}d^{n}_{m}|n,m\rangle
\;\;\;{\rm where}\;\;\;\sum_{m=1}^{k^{n}}|d^{n}_{m}|^{2}=1.       
\label{obser}
\end{equation}
At first sight Eq.~(\ref{obser}) implies that an observer 
at spatial infinity would have to attribute the BH a vanishing (Bekenstein)
entropy. However, the observer is not able to measure the coefficient 
$d_{m}^{n}$, and this lack of information corresponds to the Bekenstein
entropy.
The observable Schwarzschild Hilbert space ${\cal H}_{ob}$ is then
defined as the subspace of ${\cal H}$ which is spanned by the states defined
in Eq.~(\ref{obser}): 
\begin{equation}
{\cal H}_{ob}\equiv span\{|n\rangle_{ob}\}.
\label{obsp}
\end{equation}

We shall now explain why the above definitions are relevant to the question
discussed at the beginning of this section. The horizon splits the space in
two parts.  All the physical processes which occur in the interior part of
the BH spacetime can be viewed from the exterior only through a change in the mass
of the BH. Hence, physics inside the horizon should be described in terms of
the states 
$|n,m\rangle\in {\cal H}$, whereas the physics as it is viewed at infinity 
should
be described in terms of the observable states $|n\rangle_{ob}\in {\cal
H}_{ob}$. That is, the horizon filters out the knowledge about the internal
degrees of freedom. 
 
We shall define the horizon projection operator:
\begin{equation}
{\bf h}\equiv\sum_{n=0}^{\infty}|n\rangle\langle n|
\end{equation}
where from now on $|n\rangle\equiv|n\rangle_{ob}$. Note that the horizon
projection operator is the unit operator in the observable subspace. The
projection operator   represents the horizon in which all the parameters
describing the BH are hidden  apart from its mass. That is, each state
$|\psi\rangle$ in
${\cal H}$ and each observable ${\bf B}$ in ${\cal H}$ are reduced to
\begin{eqnarray}
|\psi\rangle & \longrightarrow & |\psi\rangle_{ob}\equiv {\bf
h}|\psi\rangle\nonumber\\ {\bf B} & \longrightarrow & {\bf B}_{ob}\equiv {\bf
h}{\bf B}{\bf h}
\label{eye}
\end{eqnarray} 
when a physical measurement is made by an observer located at large
distance from the BH. 
 
Eq.~(\ref{eye}) is a manifestation of the concept of a {\it secret} observable. It
 can be regarded as a quantum version of the ``no hair" theorems because it
implies  that an observer at infinity cannot predict the internal state of
the BH apart  from its mass, or equivalently its area (the horizon projection
operator erases the  knowledge about ${\bf G}$). There are, indeed, two
systems involved. One is the ``real" physical system  of the BH (without the
horizon filter) which is described by the mixed state given in
Eq.~(\ref{mix}). Due to the internal degrees of freedom the system has an
entropy $S=A/4{\cal L}^{2}_{p}$. The other system is  a small `slice' of the
real physical system~(\ref{obsp}), and only this `slice' is  accessible to
the observer at infinity. One can view it as if the observer is wearing  a
pair of glasses that filter out all but one color (that is, the observer
state 
$|n\rangle_{ob}$).
We now turn to find the form of the observable Hamiltonian.
 
\section{The Hamiltonian}\label{sec:hamin}

There are {\it two} properties which an observer located at large distance
from the BH can measure:
the mass of the BH (including the energy
in the medium between the BH and the observer)
and the energy flux (if there is one) coming from the
BH. Hence, we shall start this section by defining an operator
in the algebra which represents the mass of the BH.

\subsection{The Mass Operator}

In section~\ref{sec:ta} the set $\left( {\bf A},{\bf G} \right)$ had been 
chosen to be maximal in the sense that if ${\bf S}$ is another observable
which commutes with both  ${\bf A}$ and ${\bf G}$, then ${\bf S}$ is a
function of ${\bf A}$ and ${\bf G}$. Now, in the observable Hilbert space,
the above argument reduces to the following one: if ${\bf S}$ is an
observable in ${\cal H}_{ob}$ and $[{\bf A},{\bf S}]=0$ then 
${\bf S}$ is a function of ${\bf A}$. We shall assume that the mass operator 
${\bf M}$ which represents the BH mass commutes with ${\bf A}$. That is,
${\bf M}$ is a function of ${\bf A}$. 

Classically, the mass of the BH can be obtained
from its area by the well known relation $A=16\pi M^{2}$. We shall assume
that to a good approximation this relation also holds true for  the
operators, and hence the mass operator can be written as
\begin{equation}
{\bf M}\equiv\sqrt{\frac{\bf A}{16\pi}}(1+\epsilon({\bf A}))
\label{masso1}
\end{equation}
where the eigenvalues of ${\bf M}$ constitute the mass  spectrum of the BH.
The dimensionless function $\epsilon({\bf A})$ must approaches zero in the
limit of a massive BH in order to reproduce the  classical relation $A=16\pi
M^{2}$. Now, since $\epsilon({\bf A})$ is a  dimensionless function of ${\bf
A}$ it can be approximated, to first order, by some power of ${\cal
L}_{P}^{2}/{\bf A}$. Hence, its contribution to
Eq.~(\ref{masso1}) may be significant only for BHs of the order of the
Planck mass. We shall be interested in BHs not near the Planck mass, and thus
we {\it redefine} the mass operator as
\begin{equation}
{\bf M}\equiv\sqrt{\frac{\bf A}{16\pi}}
\label{masso}
\end{equation}
 
\subsection{General Form of the Hamiltonian}\label{sec:gfh} 

We shall now try to find the form of the Hamiltonian.
If we assume that it is equal to ${\bf M}$,
we can not understand  
why the hyperspin of the BH can have
eigenvalues only of the form $(k^{n}-1)/2$.
Moreover, this Hamiltonian describes a static system
which clashes with the existence of the HR.
 
Let us now list the demands from the correct Hamiltonian:\\
(1) It should be a Hermitian operator.
(2) For a very large mass the Hamiltonian of the 
system should be equal to the mass of the black
hole.
(3) It should allow transitions {\it only} between 
states with hyperspins $s$   
and  $s'$ which satisfy the condition $s'=s\pm k^{n}/2$
for some integer $n$.
(4) It does not contain ${\bf G}$.
The last demand arises because in the observer's Hilbert space the area operator is not
degenerate (see section~\ref{subsec:ohs}).  
From these requirements it follows that we can 
write the observable Hamiltonian in the form
\begin{equation}
{\bf H}={\bf M}+{\bf V}
\label{hamil}
\end{equation}
where ${\bf V}$ describes an ``interaction term" 
which approaches zero as the BH mass becomes very large.

In section~\ref{sec:algebra} we have defined three Hilbert spaces. The first
one, is the ``full" Hilbert space ${\cal H}_{f}$ which is spanned by the set
of hyperspin states $\{|s,m_{3}\rangle\}$. The second is a subspace
${\cal H}$ of the full  Hilbert space which is spanned by the set of states
$\{|n,m\rangle\}$ where $n$ and $m$ are given in Eq.~(\ref{nms}). The third
Hilbert space is a subspace ${\cal H}_{ob}$ of ${\cal H}$ which is spanned by
the observable states defined in Eq.~(\ref{obser}). In
section~\ref{subsec:ohs} we have explained why, in the eyes of an observer 
at infinity, ${\cal H}_{ob}$ is the relevant space  to work with. The
question remains, why we did have to define ${\cal H}_{ob}$ as a subspace of 
${\cal H}$ (with area spectrum $a_{n}=a_{0}n$) and not of ${\cal H}_{f}$ 
(with area spectrum $a_{s}=\log_{k}(2s+1)$)? Equivalently, why should the area
eigenvalues  be equally spaced?
The answer to the questions above lies in the third requirement on ${\bf H}$. 
Let $|s\rangle$ and $|s'\rangle$ be some states in ${\cal H}_{f}$ with
hyperspin $s$ and 
$s'$ respectively. Then, by implementing the third 
requirement on ${\bf H}$ we find that $\langle s'|{\bf H}|s\rangle=0$ unless 
$s=s'\pm k^{n^{*}}/2$ where $n^{*}$ is an integer. Thus, if the initial
state of the  BH, $|s\rangle$, belongs also to ${\cal H}$ (that
is $s=(k^{n}-1)/2$), then 
$\langle s'|{\bf H}|s\rangle=0$ unless $|s'\rangle\in {\cal H}$; hence  the
evolution of the system takes place in a subspace ${\cal H}$ of ${\cal
H}_{f}$.

On the other hand, if the initial state $|s\rangle$ does not belong to ${\cal
H}$ then $\langle s'|{\bf H}|s\rangle=0$ for any $|s'\rangle\in {\cal H}$ so
that ${\cal H}$ does not play any role in the evolution of the BH. In this
case, the area eigenvalue of the initial state $|s\rangle$ is
$a_{s}=\log_{k}(2s+1)$  which is not in general an integer,  but we can still
write $a_{s}=n+\eta$ where $n$ is an integer and $0<\eta<1$. This eigenvalue
is still consistent with Eq.~(\ref{intro}) given in the Introduction for the
area eigenvalues. Hence, we shall define a subspace
${\cal H}_{\eta}$ of
${\cal H}_{f}$ as follows:
\begin{equation}
{\cal H}_{\eta}\equiv span\{|s_{n}\equiv n+\eta,m_{3}\rangle\}
\end{equation}
where $n=0,1,2,...$ and $m_{3}=-s_{n}/2,-s_{n}/2+1,...,s_{n}/2$.   
In the same manner, the third  requirement on ${\bf H}$ implies that the time
evolution of the BH takes place  in ${\cal H}_{\eta}$. Hence, it is clear now
that the third demand on the Hamiltonian is consistent with the spectrum of
Eq.~(\ref{intro}). From now on, for simplicity,  we choose $\eta=0$ (note
that ${\cal H}={\cal H}_{\eta=0}$).    
 
We have explained why the third demand on ${\bf H}$ is
necessary in order to shift attention from the full Hilbert space to the
subspace
 ${\cal H}$ which yields the degeneracy $k^{n}$ for the number operator ${\bf
N}$. We shall now be interested in the  kind of interaction term ${\bf V}$
which implements the third requirement.  

We introduce raising 
and lowering operators ${\bf E^{\dag}}$ and ${\bf E}$ of {\bf A}, respectively, 
which we shall define explicitly later. These operators must
satisfy the following commutation relation
\begin{equation}
[{\bf A},{\bf E^{\dag}}]={\bf E^{\dag}}\;\;\; {\rm and}
\;\;\;[{\bf A},{\bf E}]=-{\bf E}. 
\label{ecr}
\end{equation} 
From the third requirement on ${\bf H}$ it is quite
clear that ${\bf V}$ contains ${\bf E^{\dag}}$ and 
${\bf E}$ since $\left({\bf E}\right)^{n}$, for example, allows only 
transitions  between states with hyperspins $s$  and $s-k^{n}/2$.
In addition, we are not interested in  products of the form ${\bf
E^{\dag}}{\bf E}$ since they keep ${\bf A}$ unchanged (that is, 
$[{\bf A},{\bf E^{\dag}}{\bf E}]=0$).
Combining all of this with the hermiticity 
of ${\bf V}$ we find that ${\bf V}$ can be written
as 
\begin{equation}
{\bf V}=h({\bf A},{\bf E})+h^{*}({\bf A},{\bf E}^{\dag})
\label{fov}
\end{equation}
where $h$ is a complex function of ${\bf A}$ and ${\bf E}$.

Now we shall expand the function $h$ as a power series in ${\bf E}$. That
is, the first term is proportional to ${\bf E}$, the second one to 
$\left({\bf E}\right)^{2}$ etc. It is reasonable to assume that in a
perturbative theory the transitions $n\rightarrow n\pm 1$ have more weight
than higher order transitions  (such as $n\rightarrow n\pm 2$). Hence  we
shall assume that up to a given approximation, $h$ is linear in ${\bf E}$.
Later on, we shall justify this assumption by  a comparison with the HR for a
massive  BH. But note that despite the linearization of $h$, the time-evolution
operator, which is an exponential function of ${\bf E}$ and ${\bf E}^{\dag}$,
contains all orders of these operators. We can thus write the ``interaction"
term  as
\begin{equation}
{\bf V}=f({\bf A}){\bf E}+{\bf E}^{\dag}f^{*}({\bf A}),
\label{iaph}
\end{equation}
where $f$ is some function of ${\bf A}$. If we normalize the 
raising and lowering operators to have expectation values
of the order of unity (as we shall see shortly), it follows from
the second requirement on the Hamiltonian that the function
$f(n)$ approaches zero as the parameter $n$ 
($n=\langle{\bf A}\rangle$) approaches infinity. The question is how fast
$f(n)$ approaches zero. We shall give here a motivation that $f(n)\sim
1/\sqrt{n}$ for large
$n$, and in section~\ref{sec:hr} we shall estimate the function $f(n)$ 
in a more deductive manner.

It was shown by Kastrup~\cite{Kast97} that there is
a small correction to the BH entropy calculated in canonical ensemble of the
order of the logarithm of the area. Likewise, we~\cite{Gour99} have
obtained  a logarithmic correction to the entropy by using the
grand-canonical approach (we have shown that the grand-canonical approach  is
also compatible with the HR).  Padmanabhan~\cite{Padi}, as well as Kim, Lee and
Ji~\cite{KLJ}, have  shown that for a small number of particles
in the background of the Schwarzschild metric,
(or if the system obeys Boltzmann statistics), the
entropy is proportional to the logarithm of the 
area. This implies that there might be a correction
to the entropy of the order of the logarithm of 
$A$. Motivated by the above arguments we shall write
the entropy as:
\begin{equation}
S=\frac{1}{4{\cal L}_{P}^{2}}A+b\ln(d A)
\label{ent}
\end{equation}
where $b$ and $d$ are constants. Thus
we find a small correction to the energy $U$ of
the BH:  
\begin{equation}
dU=T_{BH}dS=dM+\frac{b '}{M^{2}}dM
\label{fof}
\end{equation}
where $b '$ is some constant. Note that the correction to Hawking 
temperature $T_{BH}$, if there is one, is absorbed in $b '$. Hence, the
correction  to the energy $U$ is proportional to the reciprocal of $M$. This
is one reason for guessing that the function $f({\bf A})$
in Eq.~(\ref{iaph}) is proportional to $1/{\sqrt{\bf A}}$ 
for large  $\langle{\bf A}\rangle$ values. As we have mentioned earlier, 
we shall later
derive this result in a more rigorous way. We now turn to define the
operators ${\bf E}$ and ${\bf E}^{\dag}$ explicitly.

The most general form of a raising operator ${\bf a}^{\dag}$ in 
the Hilbert space ${\cal H}$ is given by
\begin{equation}
{\bf a}^{\dag}=\sum_{n=0}^{\infty}\sum_{m_{1}=1}^{k^{n}}
\sum_{m_{2}=1}^{k^{n+1}}C_{m_{1}m_{2}}^{n}|n+1,m_{2}
\rangle\langle n,m_{1}|
\label{roa}
\end{equation}
where $C_{m_{1}m_{2}}^{n}$ are complex numbers. The lowering
operator ${\bf a}$ is defined by taking the Hermitian conjugate of
Eq.~(\ref{roa}). Now, according to Eq.~(\ref{eye}), in the observable Hilbert
space ${\cal H}_{ob}$, the raising operator ${\bf a}^{\dag}$ transforms to 
${\bf h}{\bf a}^{\dag}{\bf h}$. That is, the raising operator in 
${\cal H}_{ob}$
is defined in terms of the observable states $|n\rangle$ instead of 
Eq.~(\ref{roa}). 
Hence, we are able to define ${\bf E}$ and 
${\bf E}^{\dag}$ as the Susskind-Glogower
operators~\cite{SusGlo64}, 
\begin{equation}
{\bf E}=\sum_{n=0}^{\infty}|n\rangle\langle n+1|\;\;\;{\rm and}\;\;\; 
{\bf E}^{\dag}=\sum_{n=0}^{\infty}|n+1\rangle\langle n|.
\label{defsgo}
\end{equation}
It is clear from the definition that these operators 
satisfy relation~(\ref{ecr}). We can define also the standard
creation and annihilation operators, ${\bf a}$ and ${\bf a}^{\dag}$,
respectively, as
\begin{equation}
{\bf a}\equiv ({\bf N}+1)^{1/2}{\bf E},\;\;\;
{\bf a}^{\dag}\equiv {\bf E}^{\dag}({\bf N}+1)^{1/2}.
\label{sgo}
\end{equation}
It is clear that ${\bf N}={\bf a}^{\dag}{\bf a}$, and
\begin{equation}
[{\bf a}^{\dag},{\bf a}]=\sum_{n=0}^{\infty}|n\rangle
\langle n|=1_{{\rm ob}}
\label{ccr}
\end{equation}
where the $1_{{\rm ob}}$ indicates a unit operator in ${\cal H}_{ob}$ (notice
that it is {\it not} a unit operator in ${\cal H}$). To write the interaction
term~(\ref{iaph}) with the operators ${\bf E}$ and ${\bf E}^{\dag}$ defined in
Eq.~(\ref{defsgo}) requires knowledge of the quantum phase problem. Hence,
before we continue with the formalism, let us first review this problem in
accordance with Lynch~\cite{Lyn95}.

\subsection{The Quantum Phase Problem}  

Classically, the electric field vector $\vec{E}(\vec{r},t)$, of an 
electromagnetic field contained in a cubic cavity of edge L, may be expanded
in Fourier components:
\begin{equation}
\vec{E}(\vec{r},t)\propto\sum_{\vec{k}}\left(a_{\vec{k}}{\rm e}^{i(\vec{k}
\cdot\vec{r}-\omega t)}+
a_{\vec{k}}^{*}{\rm e}^{-i(\vec{k}\cdot\vec{r}-\omega t)}\right)
\end{equation}
where we assumed for simplicity a linearly polarized field
at a single frequency $\omega$. The sum is taken over all 
the plane wave modes, where $\vec{k}$ is the wave vector.
If we are limited to a single excited mode, say, 
$\vec{k}=\vec{k}_{0}$, we can write
\begin{equation}
\vec{E}(\vec{r},t)\propto\left(a{\rm e}^{i(\vec{k_{0}}\cdot\vec{r}-\omega t)}+
a^{*}{\rm e}^{-i(\vec{k_{0}}\cdot\vec{r}-\omega t)}\right)
\end{equation}
where $a\equiv a_{\vec{k_{0}}}$. Writing $a=r{\rm e}^{i\phi}$
we finally obtain
\begin{equation}
\vec{E}(\vec{r},t)=\vec{E}_{0}\cos (\vec{k_{0}}
\cdot\vec{r}-\omega t+\phi).
\end{equation}
In this way we $decompose$ the classical electric field into
amplitude and phase components. 

In quantum mechanics 
$a\rightarrow {\bf a}$ and $a^{*}\rightarrow {\bf a}^{\dag}$
where ${\bf a}$ and ${\bf a}^{\dag}$ satisfy the canonical
commutation relation~(\ref{ccr}). 
Dirac (1927) was the first to attempt to define 
a phase operator. He decomposed ${\bf a}$ into an amplitude
and a phase component
\begin{equation}
{\bf a}={\rm e}^{i{\bf\phi}}{\bf N}^{1/2}
\label{dirac}
\end{equation}
where ${\bf N}={\bf a}^{\dag}{\bf a}$.
The commutator $[{\bf a},{\bf a}^{\dag}]=1$ leads to the Lerner criterion
\begin{equation}
[{\rm e}^{i{\bf\phi}},{\rm N}]={\rm e}^{i{\bf\phi}}.
\label{icr}
\end{equation}
Expanding ${\rm e}^{i\phi}$ on both sides 
of relation~(\ref{icr}) gives
\begin{equation}
[{\bf N},{\bf\phi}]={\rm i}
\label{npu}
\end{equation}
and thus shows that ${\bf N}$ and ${\bf\phi}$ are canonically
conjugate operators. This relation leads to the 
$number-phase$ uncertainty relation of quantum optics:
\begin{equation}
\Delta{\bf N}\Delta{\bf\phi}\geq\frac{1}{2}.
\label{aphu}
\end{equation}
That is, to create a radiation with a sharp phase, such as a laser beam, 
we would need to have many photons. Similarly, as we 
shall see later, a massive BH (analog of many quanta) has a sharp phase.  
 
But Eq.~(\ref{npu}) leads to serious problems.
For example, if we take its matrix elements  
in the number state basis, we
find that
\begin{equation}
(n-n')\langle n|{\bf\phi}|n'\rangle=i\delta_{n,n'}
\end{equation}
which leads to the impossible equation 
$0=$i for $n=n'$. Moreover, Susskind and Glogower~\cite{SusGlo64}
have shown that ${\rm e}^{i\bf\phi}$
is not unitary, or equivalently, $\bf\phi$ is not
Hermitian.

The recent paper by Bojowald, Kastrup, Schramm 
and Strobl~\cite{BKSS99} suggests a solution to these problems.
However, as we shall see shortly,
despite these problems, if one does not define the phase as an operator
(just as time is not an operator in quantum mechanics), the above 
criticism is not a decisive  objection to the phase-number uncertainty 
relation~(\ref{aphu}). For although we have no time operator, 
we have time-energy uncertainty relation. 
The reason why it is difficult to define the phase or the time
operator is the fact that the  particle number or energy spectrum is
non-negative and, therefore, is bounded from below. This situation is
drastically different from the one for position and momentum whose eigenvalue
spectra are unbounded.  

Let us now compare the Dirac 
definition for the phase operator 
Eq.(\ref{dirac}), and Eq.~(\ref{sgo}). One
can see that ${\bf E}$ and ${\bf E}^{\dag}$
are analogous of ${\rm e}^{i{\bf\phi}}$
and ${\rm e}^{-i{\bf\phi}}$ respectively.
Now, using the definition Eq.~(\ref{defsgo})
it is easy to show that ${\bf E}$ and 
${\bf E}^{\dag}$ are ``one-sided unitary" or an 
$isometry$. That is,
\begin{equation}
{\bf E}{\bf E}^{\dag}=1\;\;\;{\rm but}\;\;\;
{\bf E}^{\dag}{\bf E}=1-|0\rangle\langle 0|.  
\end{equation}
Despite the fact that the vacuum projection 
``spoils" the unitarity of ${\bf E}$, for
states with a very large number occupations
(that is, $\langle {\bf N}\rangle\gg 1$) and
hence small vacuum component, we can treat ${\bf E}$
as an  approximately unitary operator. As we shall see now, the last statement
helps us to define the {\it phase} of a BH.

\subsection{The Phase of a Black Hole}

For a BH not near the Planck mass, 
$\langle {\bf N}\rangle\gg 1$. Thus we 
can define a phase operator by
\begin{equation}
{\rm e}^{i{\bf \phi}}\approx {\bf E}.
\label{defph}
\end{equation}
Hence, the interaction term ${\bf V}$, which
has the form of Eq.~(\ref{iaph}), describes 
an interaction between the area operator ${\bf A}$
and the phase operator ${\bf\phi}$. As we shall
see later, the HR in our model 
is due to an area-phase interaction.

The definition of the quantum phase of a BH in Eq.~(\ref{defph}) raises the
question, what is the physical interpretation of the ``new" observable
${\bf\phi}$  ?   Furthermore, what are the physical grounds for an area-phase
interaction appearing in the Hamiltonian?  In order to answer these
questions, we first note that the phase operator is the canonical conjugate
of the area operator (see Eq.~(\ref{npu})). This is  reminiscent of work by
Carlip and Teitelboim~\cite{CarTe95}.

Carlip and Teitelboim have shown by using the standard  (Euclidean) action
principle for gravitational field, that the horizon area and the  opening
angle at the horizon are canonical conjugates.  They start with the wave
functional $\psi=\psi[T,^{(3)}g]$ where the separation time at spatial
infinity $T$, is an extra argument (in addition to the intrinsic geometry of
a spatial section $^{(3)}g$) since space is not closed. The canonical
conjugation relation of $T$ and the ADM mass $M$ is expressed in the 
Schr\"odinger equation
\begin{equation}
\frac{\hbar}{i}\frac{\partial\psi}{\partial T}+M\psi=0
\end{equation}
which can be obtained from an extended Wheeler-DeWitt equation (that is, the
appropriate action is the canonical action $S_{m+g}$ supplemented by $-TM$).
In the same manner they introduce an extra parameter $\Theta$
(``dimensionless internal time")  associated with the horizon by the relation
\begin{equation}
\frac{\hbar}{i}\frac{\partial\psi}{\partial\Theta}-A\psi=0
\end{equation}
where $A$ is the horizon area. They 
work in a system of polar coordinates $r$,$\tau$ in $\Re^{2}$ (the spacetime
topology is $\Re^{2}\times S^{d-2}$) where they set the origin $r=r_{+}$ to be the 
horizon of the BH and the angle $\tau$ to be the Killing time (for more
details  see Carlip and Teitelboim paper~\cite{CarTe95}).  Next, they show
that the parameter $i\Theta$ is `` the total proper angle (proper length
divided by proper radius) of an arc of very small radius and coordinate
angular opening $\tau_{1}-\tau_{2}$".

The main result of Carlip and Teitelboim is that the canonical conjugate of the 
horizon area is the opening angle. The Euclidean continuation of the
hyperbolic  angle $\Theta$ is bounded and even periodic. This not only favors
our simple model,  but it gives a motivation for the assumption that the
horizon area (in the Euclidean sector) can be represented by a {\it number}
operator~! We are able to define the phase in Eq.~(\ref{defph}) since we have
assumed that the area spectrum is equally spaced. If, for example, the area
spacing $a_{n+1}-a_{n}$ approached zero for large quantum  number~$n$ (as
suggested by Ashtekar and others~\cite{AshKra}),  the quantity canonically
conjugate  to area would not be a phase operator (its eigenvalues would not
be bounded). 
 
In our simple model, the canonical conjugate of the horizon area is also some
kind of an angle. Since in our model the phase (angle) is an operator, we
shall identify the eigenvalues of ${\bf\phi}$ with the ``opening angles" at
the horizon, as discussed by Carlip and Teitelboim. As we shall see below, a
further analogy between our approach  and theirs can be found in the
description of the HR.

Recently, Massar and Parentani~\cite{MasPar99} have shown that the probability for a BH 
to emit a particle is given by 
\begin{equation}
P_{M\rightarrow M-\lambda}=N(\lambda,M){\rm e}^{-\Delta A(\lambda,M)/4} 
\label{imha}
\end{equation}
where $\Delta A$ is the change in the horizon area
induced by the emission, and $N$ is some phase space factor. 
Since they did not neglect the specific heat of the BH,
expression~(\ref{imha}) improves Hawking's result. We shall now point out the origin
of this result.

As suggested by Carlip and Teitelboim, Massar and Parentani have supplemented the 
canonical action with two boundary terms, 
$-TM$ (boundary at infinity) and $\Theta A/8\pi$
(boundary at the horizon). Hence, they face four different actions  
according to which quantities are fixed in the variational principle.
They fix the ADM mass $M$ and the opening angle $\Theta$ in order to
describe a physical situation in which the BH and the surrounding matter
 exchange energy, but no energy is exchanged with infinity (this is also the
process we
 are  studying). It is through the boundary term $\Theta A/8\pi$ that they
obtain Eq.~(\ref{imha}).  

In our approach, as we shall see later, the HR appears as a result of the
interaction~(\ref{iaph}), an interaction between 
the phase (``opening angle") and the area of the BH. And Massar and 
Parentani have obtained Eq.~(\ref{imha}) after introducing the boundary term 
$\Theta A/8\pi$ in the action; this can be considered (in the context of our 
model) as a `coupling' between the horizon area and the opening angle.  Hence
the Massar-Parentani approach is compatible with our model. For this reason we
identify the area-phase interaction term in Eq.~(\ref{iaph}) with
the boundary term $\Theta A/8\pi$.   Note that in our approach the
interaction term appears in the  Hamiltonian, whereas in the Massar-Parentani
approach the boundary term appears in the  action. Furthermore, it is
interesting to note that the effect of general relativity (that is, curved spacetime) 
appears in our simple model  as a coupling between the 
observable ${\bf A}$ and its {\it own} canonical conjugate ${\bf\phi}$~! 

We now ask what our model has to say about radiation by the BH. In addition to the 
Hamiltonian we need specify the state of the BH.

\section{The Initial State}\label{sec:tis} 
 
One might think that the
initial state is just the an area eigenstate $|n_{0}\rangle$ where 
$M=m_{0}\sqrt{n_{0}}$ is the initial mass of the BH. But as we shall see in 
section~\ref{sec:sbh}, an area eigenstate describes a static BH 
(that is, no HR).
The question is then, what is the initial state which leads to the 
semi-classical radiation (HR), and why an area eigenstate is not an appropriate
candidate.

The question above is not unique to our problem, but is a  general question
in quantum mechanics: What are the states that most closely imitate classical
behavior? The simplest case where this question appears is that of the 
simple harmonic oscillator. Hence we shall first review, very shortly, what
are the states that imitates the classical harmonic oscillations (see
Sakurai). 

In the Heisenberg picture the position operator ${\bf x}(t)$ 
of an harmonic oscillator is given by
\begin{equation}
{\bf x}(t)={\bf x}(0)\cos\omega t+\left(\frac{{\bf p}(0)}{m\omega}\right)
\sin\omega t
\end{equation}  
where $\omega$ is the angular frequency of the classical oscillator
and ${\bf p}(0)$ is the momentum operator at $t=0$. The expectation value
$\langle n|{\bf x}(t)|n\rangle$ vanishes because the operators ${\bf x}(0)$
and ${\bf p}(0)$ change $n$ by $\pm 1$ and $|n\rangle$ and $|n+1\rangle$ are
orthogonal. This result is different from the classical result no matter how
large 
$n$ may be. In order to obtain the classical result, we have to
take the expectation value with respect to a coherent state.

Let us perform the Bogoliubov transformation for coherent states~\cite{Umeza}:
\begin{equation}
{\bf a}\rightarrow {\bf\alpha (\lambda)}={\bf a}-\lambda
\label{Bog}
\end{equation}
where $\lambda$ is a c-number. Then, a coherent state is defined as the vacuum for
${\bf\alpha (\lambda)}$, and is denoted by $|0(\lambda)\rangle$:
\begin{equation}
{\bf\alpha (\lambda)}|0(\lambda)\rangle=0 \;\; {\rm or}\;\; 
{\bf a}|0(\lambda)\rangle = \lambda|0(\lambda)\rangle.
\end{equation}
We then have
\begin{equation}
{\bf\alpha (\lambda)}={\bf U}_{c}(\lambda){\bf a}{\bf U}^{-1}_{c}(\lambda)
\;\;\;{\rm with}\;\;\; {\bf U}_{c}(\lambda)=\exp\left(i{\bf G}_{c}(\lambda)
\right)
\end{equation}
in which the generator ${\bf G}_{c}(\lambda)$ is given by:
\begin{equation}
{\bf G}_{c}(\lambda)=i(\lambda^{*}{\bf a}-{\bf a}^{\dag}\lambda).
\end{equation}
We note parenthetically that if we allow $\lambda$ to be a function of 
${\bf A}$, we shall be able
to express the interaction term in Eq.~(\ref{iaph}) as the generator of BH 
coherent  states. That is, 
\begin{equation}
{\bf V}={\bf G}_{c}\left(\lambda({\bf A})\right)
\;\;\; {\rm with}\;\;\;\lambda({\bf A})=if^{*}({\bf A})({\bf N}+1)^{-1/2}.
\end{equation}
This remarkable result implies that the interaction term given in
Eq.~(\ref{iaph}) induces the Bogoliubov transformation~(\ref{Bog}); we shall
see that it defines the BH states which exhibit HR.  

Let us now list two properties of a coherent state which will be relevant to
our problem. Firstly, a coherent state may be expressed as a superposition of
the number (${\bf N}$) eigenstates
\begin{equation}
|0(\lambda)\rangle={\bf
U}_{c}(\lambda)|0\rangle=\sum_{n=0}^{\infty}q(n)|n\rangle
\end{equation}
where the distribution of $|q(n)|^{2}$ with respect to $n$ is of the Poisson 
type
about some mean value $\bar{n}=|\lambda|^{2}$:
\begin{equation}
|q(n)|^{2}=\left(\frac{\bar{n}^{n}}{n!}\right)\exp (-\bar{n}).
\label{poison}
\end{equation}
Secondly, a coherent state satisfies the minimum uncertainty product relation.

By analogy with a simple harmonic oscillator, we shall ask how can 
we construct a superposition of the BH area eigenstates that
most closely imitates the semi-classical result~(HR)? In comparison
with the harmonic oscillator, would it be some kind of a coherent state?
We shall call the state that does the desired job 
the coherent-termal state\footnote{The termal feature of this state has 
been actually erased after performing the projection operation of the horizon.}
and denote it by $|CT;n_{0}\rangle$ where 
$n_{0}\equiv\langle CT;n_{0}|{\bf N}|CT;n_{0}\rangle$.
Hence, in the limit of a massive BH we demand
\begin{equation}
\lim_{n_{0}\gg 1}\langle CT;n_{0}|{\bf M}(t)|CT;n_{0}\rangle
=\langle M(t)\rangle
\end{equation}
where $\langle M(t)\rangle$ is the result obtained in the semi-classical 
approach (see section~\ref{sec:hr}).

An observer at infinity is limited by the energy-time (area-phase)
uncertainty principle when he tries to determine the mass (area) of a BH
formed by a gravitational collapse. If the BH is an eternal one, in which one
can determine exactly its mass, there will be no HR because the initial state
will be a mass  eigenstate (see section~\ref{sec:sbh}). By contrast, if the
BH has been formed by gravitational collapse, there is always radiation in
the region between the BH and the observer. Even if hypothetically at some
initial time, say $t_{0}$, there is no  radiation, still the measurement
takes time, say $\Delta t$, in which  a small amount of radiation is produced
according to the mass-time uncertainty principle. Hence, in order to be compatible with
the HR, the initial state must satisfy  $\langle (\Delta {\bf
M})^{2}\rangle\ne 0$, or equivalently, the HR is due to the  uncertainty (of
an observer at infinity) about the area (mass) of the BH. 
 
The coherent-thermal state (CT-state) may be expressed as follows:
\begin{equation}
|CT;n_{0}\rangle=\sum_{n=0}^{\infty}P_{n}^{1/2}\exp(-i\nu (n))|n\rangle
\end{equation}
where $P_{n}$ is the distribution with respect to $n$
and $\exp(-i\nu (n))$ is some phase coefficient where $\nu (n)$ is a real
function of $n$. We can set $\exp(-i\nu (n))=1$
by defining $|n\rangle_{\rm new}\equiv \exp(-i\nu (n))|n\rangle$ and doing
all the previous analysis with respect to these new states. The new
Susskind-Glogower's  operators, ${\bf E}$ and ${\bf E}^{\dag}$, are different
from the old ones, but  the form of the Hamiltonian and the interaction term
given in Eq.~(\ref{iaph}) are  unchanged since $f({\bf A})$ can absorb the
changes. Hence we conclude that the  CT-state may be written as:
\begin{equation}
|CT;\bar{n}\rangle=\sum_{n=0}^{\infty}P_{n}^{1/2}|n\rangle.
\end{equation}  

We are interested in the semi-classical limit $n_{0}\rightarrow\infty$. In
this limit $\sigma_{M}/\langle {\bf M}\rangle\rightarrow 0$ where 
$\sigma_{M}^{2}\equiv \langle M^{2}\rangle-\langle M\rangle^{2}$. 
Hence, in the limit of large $n_{0}$ value, the distribution $P_{n}$ 
is of the Gaussian type about the mean value $n_{0}$~\cite{Beke84}. That is,
\begin{equation}      
P_{n}\approx\left(\frac{\alpha}{4\pi n_{0}}\right)
^{\frac{1}{2}}\exp\left(-\frac{\alpha}{4n_{0}}(n-n_{0})^{2}\right)
\end{equation}
where we have left the variance, $\sigma_{A}^{2}=2n_{0}/\alpha$, to be 
determined later. But a distribution of the Poisson type (see
Eq.~(\ref{poison})) becomes, in the limit $n_{0}\rightarrow\infty$,
Gaussian   with a variance $\sigma_{A}^{2}=n_{0}$. Thus, for a coherent state
$\alpha=2$. However, to be more general we shall not yet set $\alpha=2$.

Let us now compare the semi-classical results and 
our quantum mechanical approach when the averages are taken 
with respect to the CT-state 
\begin{equation}
|CT;n_{0}\rangle=\left(\frac{\alpha}{4\pi n_{0}}\right)^{\frac{1}{4}}\sum_{n=0}
^{\infty}\exp\left(-\frac{\alpha}{8n_{0}}(n-n_{0})^{2}\right)|n\rangle.
\label{thst}
\end{equation}

\section{The Role of Hawking Radiation}\label{sec:hr}

We start by emphasizing the similarity 
of the following two roads to the HR:\\
(1)If one attributes to a BH an entropy
$S_{BH}=A/4{\cal L}_{P}^{2}$ (Bekenstein~\cite{Bek73}), 
it leads to a temperature 
$T_{BH}=\hbar/8\pi M$ (Hawking~\cite{Hawk75});
hence the BH radiates and decreases
its horizon area ($A=16\pi M^{2}$). Due to the thermal character of the 
radiation, one can calculate the rate of change of the BH mass using
Stefan-Boltzmann law (see Eq.~(\ref{avom})).\\ 
(2)BH entropy, together with the assumption 
that the horizon area is represented by the number operator
${\bf A}=4{\rm ln}k{\cal L}_{P}^{2}{\bf N}$, implies
a huge degeneracy with respect to ${\bf N}$, $k^{n}$. This degeneracy 
motivated us to introduce into the Hamiltonian an interaction
term of the form of Eq.~(\ref{fov}). A BH with such high degeneracy
acts in order to reduce its internal degrees of freedom. Hence the area
of the BH is not conserved, i.e. 
$[{\bf H},{\bf A}]\ne 0$.
Later on we shall see that it is indeed possible to choose
the interaction term such that the area of the BH decreases in time.

The purpose of this section is to compare these two roads. Such a comparison 
 examines the validity of our assumptions that lead to the form of the 
Hamiltonian given in the previous section. Moreover, we shall be able to
derive the form of the function $f({\bf A})$ in Eq.~(\ref{iaph}) for large
$\langle {\bf A}\rangle$ values. We shall point out that these two roads
should match {\it only} in the limit of large $\langle {\bf A}\rangle$ values
(the semi-classical limit). For much smaller values of $\langle {\bf
A}\rangle$ (near the Planck scale), our approach predicts new results.

Due to Hawking radiance and its thermal character, it
is possible to estimate the rate of change of the mass 
$\langle M\rangle$ of the BH. Using the Stefan-Boltzmann
law for a surface area $16\pi\langle M\rangle ^{2}$ at 
temperature $T_{BH}=(\hbar/8\pi)\langle M\rangle ^{-1}$,
we find (see Bekenstein and Mukhanov~\cite{BekMu95} and Page~\cite{Pag76})
\begin{equation}
\frac{d\langle M\rangle}{dt}=
-\frac{\gamma\hbar}{15360\pi\langle M\rangle ^{2}}
\equiv -\frac{\varepsilon}{\langle M\rangle ^{2}}
\label{avom}
\end{equation}
where $\gamma$ lumps uncertainties about the species of particles emitted.
We shall assume that the mass of the BH
at time $t=0$ is equal to the total energy of the system, that is, $\langle
M(0)\rangle$ is the ADM mass. Now, solving Eq.~(\ref{avom}) we find
\begin{equation}
\langle M(t)\rangle=\langle M(0)\rangle\left(1-\frac{3\varepsilon t} {\langle
M(0) \rangle ^{3}}\right)^{1/3}=\langle M(0) \rangle-\frac{\varepsilon}
{\langle M(0) \rangle ^{2}}t-\frac{\varepsilon ^{2}} {\langle M(0) \rangle
^{5}}t^{2}-\cdots
\label{taom}
\end{equation}
where $\varepsilon$ is defined in Eq.~(\ref{avom}) and
the mass of the BH is assumed to be very large
compared to $(\varepsilon t)^{3}$.

In our model, the mass of the BH is represented by the mass operator ${\bf
M}\equiv {\bf M}_{S}\equiv {\bf M}_{H}(0)$ where the subscripts $S$ and $H$
stand for the Schr\"odinger and Heisenberg pictures,  respectively. The
evolution in time of the mass operator in the Heisenberg picture is
determined by
\begin{eqnarray}
{\bf M}_{H}(t) & = & \exp\left(\frac{i}{\hbar}{\bf H}t\right)
{\bf M}(0)\exp\left(-\frac{i}{\hbar}{\bf H}t\right)
={\bf M}(0)+\frac{i}{\hbar}[{\bf H},{\bf M}(0)]t\nonumber\\
 & + & \frac{1}{2!}\left(\frac{i}{\hbar}\right)^{2}
[{\bf H},[{\bf H},{\bf M}(0)]]t^{2}+\cdots
\label{teom}
\end{eqnarray}
where the Hamiltonian is given by Eq.~(\ref{hamil}) and Eq.~(\ref{iaph}).
In order to compare Eq.~(\ref{teom}) with the semi-classical 
result~(\ref{taom}), we have to take the average of both
sides of Eq.~(\ref{teom}). We shall first compare between the coefficients
that are proportional to $t$ in both Eq.~(\ref{taom}) and (the mean value
of)  Eq.~(\ref{teom}).

\subsection{The First Order Corrections}\label{sec:cfoc}

The Hamiltonian~(\ref{hamil}) may be written as
\begin{equation}
{\bf H}={\bf M}(0)+\sum_{n=0}^{\infty}\left(f(n)|n\rangle\langle n+1|+f^{*}(n)
|n+1\rangle\langle n|\right)
\label{newhamil}
\end{equation}
where we used Eq.~(\ref{iaph}) for ${\bf V}$. Hence it clear that
\begin{eqnarray}
& & [{\bf H},{\bf M}(0)]=[{\bf V},{\bf M}(0)]=\nonumber\\
& & m_{0}\sum_{n=0}^{\infty}\left(
(\sqrt{n+1}-\sqrt{n})\left(f(n)|n\rangle
\langle n+1|-f^{*}(n)|n+1\rangle\langle n|\right)\right).
\end{eqnarray}
Note that if we take the average with 
respect to an area eigenstate $|n\rangle$, we would find that
\begin{equation}
\langle n|[{\bf H},{\bf M}(0)]|n\rangle=0,
\end{equation}
and it would be impossible to compare with Eq.~(\ref{taom}).

Now, taking the average with respect to the CT-state~(\ref{thst}) we find
\begin{eqnarray}
& & \langle [{\bf H},{\bf M}(0)]\rangle _{CT;n_{0}}=2m_{0}
\sum_{n=0}^{\infty}f(n)
(\sqrt{n+1}-\sqrt{n})\langle |n\rangle\langle n+1|\rangle _{CT;n_{0}}=
\nonumber\\
& & 2m_{0}\left(\frac{\alpha}{4\pi n_{0}}\right)^{\frac{1}{2}} 
\sum_{n=0}^{\infty}f(n)
(\sqrt{n+1}-\sqrt{n}){\rm e}^{\left(-\frac{\alpha}{8n_{0}}(n+1-n_{0})^{2}
-\frac{\alpha}{8n_{0}}(n-n_{0})^{2}\right)}\nonumber\\
& & =2m_{0}{\rm e}^{-\frac{\alpha}{16n_{0}}}
\left(\frac{\alpha}{4\pi n_{0}}\right)^{\frac{1}{2}}
\sum_{n=0}^{\infty}f(n)(\sqrt{n+1}-\sqrt{n})
{\rm e}^{\left(-\frac{\alpha}{4n_{0}}(n-n_{0}
+\frac{1}{2})^{2}\right)}\nonumber\\
& & \stackrel{n_{0}\gg 1}{\approx}
2m_{0}f(n_{0})(\sqrt{n_{0}+1}-\sqrt{n_{0}}){\rm e}^{\left(-\frac{\alpha}
{16n_{0}}\right)}
\label{tank}
\end{eqnarray}
where we have assumed that $f(n)$ is purely imaginary since we want that 
$\langle {\bf H}\rangle _{CT;n_{0}}=\langle {\bf M}(0)\rangle _{CT;n_{0}}$
(the mass of the BH at time $t=0$ is equal to the total energy 
of the system). Now, comparing i$\hbar^{-1}\langle [{\bf H},{\bf M}(0)]
\rangle _{CT;n_{0}}$
with the coefficient of $t$ in Eq.~(\ref{taom}) we find
\begin{equation}
f(n_{0})\stackrel{n_{0}\gg 1}{=}\frac{i\hbar\varepsilon}
{m_{0}^{2}\langle M(0) \rangle}{\rm e}
^{\left(\frac{\alpha}{16n_{0}}\right)}
\label{solof}
\end{equation}
where we have taken $\sqrt{n_{0}+1}-\sqrt{n_{0}}\approx 1/(2\sqrt{n_{0}})$ 
since $n_{0}$
is very large. 
Because $\langle M(0) \rangle =m_{0}\sqrt{n_{0}}$ we conclude
\begin{equation}
f(n)\stackrel{n\gg 1}{=}\frac{i\hbar\varepsilon}
{m_{0}^{3}}{\rm e}^{\left(\frac{\alpha}{16n_{0}}\right)}\frac{1}{\sqrt{n}}.
\label{lll}
\end{equation}
Note that the $n_{0}$ in the exponent was not replaced by $n$ because this
exponent originally came from the initial state (see Eq.~(\ref{tank}) before taking
the limit of large $n_{0}$).
This equation holds only for $n\gg 1$ and there are corrections
in the limit of small $n$. Note that this result is in harmony with our previous 
estimate of $f({\bf A})$ as proportional to $1/\sqrt{\bf A}$ (see the
arguments below Eq.~(\ref{fof})).

It is clear that the function $f(n)$ should not be dependent on the
CT-state characterized by the number $n_{0}$ (otherwise, the Hamiltonian would
be dependent  on the states we take average with). Hence, the factor 
$\exp\left(\frac{\alpha}{16n_{0}}\right)$ should be independent of $n_{0}$.
There are two ways to achieve this. One is to assume that $\alpha$ is
proportional to $n_{0}$ such that the ratio
$\alpha/n_{0}$ is independent of $n_{0}$. The other way is to assume that
$\alpha$ is of the order  of unity, and since our calculations are valid up
to corrections  of O($1/n_{0}$), we can replace
$\exp\left(\frac{\alpha}{16n_{0}}\right)$ by one. By contrast, it would be
incorrect to replace $\exp\left(\frac{\alpha}{16n_{0}}\right)$  by unity if,
for example, $\alpha\sim\sqrt{n_{0}}$.  The reason is that we cannot neglect
 corrections of O($1/\sqrt{n_{0}}$) and thus the  function $f(n)$ will be 
dependent on $n_{0}$ by a small correction ($\sim1/\sqrt{n_{0}}$) which {\it
cannot} be neglected. Since the Hamiltonian should not depend on
$n_{0}$ (even not as a small correction), we conclude that either $\alpha$ is
of the order of unity, or  it is of the order of $n_{0}$. The question now is
which one of the two possibilities for $\alpha$ is the correct one.     

One may ask if the area-phase uncertainty relation~(\ref{aphu}) can supply the
required information about~$\alpha$. As we have mentioned earlier, a coherent
state (e.g.  the initial CT- state with $\alpha=2$) satisfies the minimum
uncertainty  product relation. In Appendix C, we show that both of the
choices for $\alpha$ lead to the minimum uncertainty product
relation~(\ref{aphu}). Thus we are left with an unknown parameter.

As we shall 
see shortly, Eq.~(\ref{lll}) leads to a second order correction of the form
$\sim t^{2}/\langle M(0) \rangle ^{5}$ for the evaporation of the black hole 
[see Eq.~(\ref{taom})].
This result is, surprisingly, consistent with the semi-classical  result
Eq.~(\ref{taom}). Hence, we shall compare also the coefficients of second order in both 
approaches, and deduce restrictions on $\alpha$.

\subsection{The Second Order Corrections}  

In order to calculate the second order of $\langle{\bf M}(t)\rangle$
with respect to the time $t$, we use Eq.~(\ref{teom}) and calculate
the average of
\begin{equation}
[{\bf H},[{\bf H},{\bf M}(0)]]=[{\bf M}(0),[{\bf V},{\bf M}(0)]]
+[{\bf V},[{\bf V},{\bf M}(0)]]
\label{sason}
\end{equation}
with respect to the CT-state. Due to the identity
$\langle|n\rangle\langle n+1|\rangle=\langle|n+1\rangle\langle n|\rangle$,
the first term in Eq.~(\ref{sason})
averages out to zero. To calculate the
{\it average} of the second term in Eq.~(\ref{sason}), we 
substitute Eq.~(\ref{solof}) 
in the expression for ${\bf V}$. Thus, we find
\begin{eqnarray}
\langle[{\bf V},[{\bf V},{\bf M}(0)]]\rangle_{CT;n_{0}} & = & \frac{\hbar^{2}
\varepsilon^{2}}{m_{0}^{5}}{\rm e}^{(\frac{\alpha}{8n_{0}})}
\sum_{n>1}^{\infty}\biggl\{\left(\frac{1}{n^{3/2}}-\frac{1}{(n+1)^{3/2}}\right)
\langle|n\rangle\langle n|\rangle_{CT;n_{0}}\nonumber\\ 
& & +\left(\frac{1}{n\sqrt{n+1}}
-\frac{1}{(n+1)\sqrt{n}}\right)\langle|n+2\rangle\langle
n|\rangle_{CT;n_{0}}\biggl\}
\nonumber\\
& = & \frac{\hbar^{2}\varepsilon^{2}}{2\langle M(0)\rangle ^{5}}{\rm e}^
{(\frac{\alpha}{8n_{0}})}\left(3+\sum_{n=0}^{\infty}\langle|n+2\rangle
\langle n|\rangle_{CT;n_{0}}\right)\nonumber\\
& = &\frac{\hbar^{2}\varepsilon^{2}}{2\langle M(0)\rangle ^{5}}\left(3{\rm e}^{(
\frac{\alpha}{8n_{0}})}+{\rm e}^{(-\frac{\alpha}{8n_{0}})}\right)  
\label{kolsa}
\end{eqnarray}           
where we have used the fact that the contribution of small $n$'s (compared to 
$n_{0}$) may be neglected under the average. Finally, comparing
Eq.~(\ref{kolsa}) with the semi-classical result~(\ref{taom}) we find
\begin{equation}
3{\rm e}^{(\frac{\alpha}{8n_{0}})}+{\rm e}^{(-\frac{\alpha}{8n_{0}})}=4.
\label{devi}
\end{equation}
Note that Eq.~(\ref{devi}) has two solutions. One is 
$\exp(\alpha/8n_{0})=1$. This solution is plausible, up to a very good 
approximation, if $\alpha$ is of order of unity since then
$\exp(\alpha/8n_{0})=1+{\rm O}(1/n_{0})$. A coherent state, that is
$\alpha=2$, also satisfies Eq.~(\ref{devi}).  This remarkable result shows
that our model for the Stefan-Boltzmann law agrees with the semi-classical 
result up to 
the second order in time. We have used the first order  correction to obtain
$f(n)$, and this allowed us to reproduce exactly the second order correction
(including the non-dimensional numerical coefficient). 

The other solution of Eq.~(\ref{devi}) is $\exp(\alpha/8n_{0})=\frac{1}{3}$ 
which is  impossible for $\alpha>0$.
Hence, it seems at first sight that a solution for $\alpha$ of 
order $n_{0}$ is impossible. 

In the last argument we have assumed that the 
interaction term in Eq.~(\ref{iaph}) is exact. If one 
insists that the initial CT-state, which imitates the semi-classical result,
 is such that $\alpha\sim n_{0}$ (and not of the order of unity), we must
then add a  correction to the Hamiltonian. In appendix B we have  estimated
the second order correction (if there is one) to the Hamiltonian, and we have
reproduced Eq.~(\ref{devi}) with a correction coming from the extra term in
the Hamiltonian.  We shall emphasize here that the
second order correction contributes Eq.~(\ref{devi})
because we have calculated $\langle {\bf M}(t)\rangle$ up to 
a {\it second} order in time.  

\section{Static Black Holes}\label{sec:sbh}

At first glance it seems impossible to describe a static BH in our model
since $[{\bf H},{\bf A}]\ne 0$. But, as we have seen in
section~\ref{sec:cfoc}, the first  order correction to the rate of change of
the mass of the BH is zero when the  initial state is taken to be an area
eigenstate. Thus, area eigenstates seem to be  good candidates for describing
eternal BHs. Note that these states satisfy
$\langle (\Delta {\bf M})^{2}\rangle=0$ which is compatible with eternal BHs
since no gravitational collapse is involved (see section~\ref{sec:tis}). As
we shall see in this section, for a BH (not a primordial one)  which is
described by an area eigenstate, we have to wait  a long time (much greater
than the age of the universe) in order to observe any change  in its horizon
area, and even then it will only be a small fluctuation about the  mean area
value.

The idea that area eigenstates describe static BHs has interesting 
implications.
For example, if one had a `device' able to measure the area of the BH 
(that is, the BH state collapse into an area eigenstate), it would, 
by measuring it, stop the 
time evolution of the BH! This means that our knowledge about the 
area of the BH plays a central role, and {\it affects} its mass evaporation.     
 
We start with the Hamiltonian given in Eq.~(\ref{newhamil}) where $f(n)$ is 
given by  Eq.~(\ref{lll}) for large $n$. As we have done in the previous
sections, since  $f(n)$ approaches zero for large $n$ values, we shall treat
the interaction term in  a perturbative manner. This time we shall work in
the interaction picture.

The {\it time dependent} ``potential" in the interaction picture is given by
\begin{equation}
{\bf V}_{I}(t)=\exp\left(\frac{i}{\hbar}{\bf M}(0)t\right){\bf V}
\exp\left(-\frac{i}{\hbar}{\bf M}(0)t\right)
\label{ini}
\end{equation}
where ${\bf V}$ is the interaction term given in Eq.~(\ref{newhamil}). The 
time
evolution operator (in the interaction picture) is then expressed as
\begin{eqnarray}
{\bf U}_{I}(t) & = & {\cal T}\;\exp\left(-i\hbar^{-1}\int_{0}^{t}{\bf V}_{I}
(t')dt' \right)\nonumber\\
& = & 1-\frac{i}{\hbar}\int_{0}^{t}{\bf V}_{I}(t')dt'+(-\frac{i}{\hbar})^{2}
\int_{0}^{t} dt'\int_{0}^{t'}dt'' {\bf V}_{I}(t'){\bf
V}_{I}(t'')+\cdots\nonumber\\
\label{teo}
\end{eqnarray}  
where ${\cal T}$ is the time ordering operator. We proceed to evaluate the 
average of the area operator when the initial state is the area 
eigenstate $|n_{0}\rangle$.
We shall calculate the average up to the first non vanishing order. That is,
\begin{eqnarray}
\langle {\bf A}\rangle_{t} & = & \langle n_{0}|{\bf U}^{\dag}_{I}(t){\bf A}
{\bf U}_{I}(t)
|n_{0}\rangle=n_{0}a_{0}\nonumber\\
& + & \frac{1}{\hbar^{2}}\langle n_{0}|\int_{0}^{t}{\bf V}_{I}(t')dt'
\;{\bf A}\int_{0}^{t}
dt'' {\bf V}_{I}(t'')|n_{0}\rangle\nonumber\\
& - & \frac{n_{0}a_{0}}{\hbar^{2}}\langle n_{0}|\int_{0}^{t}dt'\int_{0}^{t'}
dt'' {\bf V}_{I}(t'){\bf V}_{I}(t'')|n_{0}\rangle+{\rm O}\left({\bf
V}_{I}^{4}\right)
\label{avofa}
\end{eqnarray}
where we have used Eq.~(\ref{teo}) (note that the area operator in the 
interaction  picture is the same as in Schr\"odinger picture). 

Now, using Eq.~(\ref{ini}) and 
Eq.~(\ref{newhamil}) to express ${\bf V}_{I}(t)$, we find after some algebraic 
manipulations that Eq.~(\ref{avofa}) can be written as
\begin{eqnarray} 
\langle {\bf A}\rangle_{t} & = & n_{0}a_{0}\nonumber\\
& - & \frac{4a_{0}}{\hbar^{2}}\left(\frac{|f(n_{0}-1)|^{2}}{\omega^{2}
_{n_{0},n_{0}-1}}\sin^{2}\left(\frac{1}{2}\omega_{n_{0},n_{0}-1}t\right)-
\frac{|f(n_{0}
)|^{2}}{\omega^{2}_{n_{0},n_{0}+1}}\sin^{2}\left(\frac{1}{2}
\omega_{n_{0},n_{0}+1}t
\right)\right)\nonumber\\
\label{aita}
\end{eqnarray}
where
\begin{equation}
\omega_{n_{0},n_{0}\pm 1}\equiv\frac{E_{n_{0}}-E_{n_{0}\pm 1}}{\hbar}\equiv
\frac{m_{0}
\sqrt{n_{0}}-m_{0}\sqrt{n_{0}\pm 1}}{\hbar}.
\end{equation}
Now, because $f(n_{0})$, and $\omega_{n_{0},n_{0}\pm 1}$ are of 
O($1/\sqrt{n_{0}}$) for $n_{0}\gg 1$, we find that
$|f(n_{0}-1)|^{2}/\omega^{2} _{n_{0},n_{0}-1}$ is equal to
$|f(n_{0})|^{2}/\omega^{2}_{n_{0},n_{0}+1}$ up to a very small correction of
the order of $1/n_{0}$. Thus, after using Eq.~(\ref{lll}) for $f(n)$ we find 
\begin{equation}
\frac{|f(n_{0}-1)|^{2}}{\omega^{2}_{n_{0},n_{0}-1}}=
\frac{|f(n_{0})|^{2}}{\omega^{2}_{n_{0},n_{0}+1}}=\frac{4\hbar^{4}
\varepsilon^{2}} {m_{0}^{8}}\exp\left(\frac{\alpha}{8n_{0}}\right)
\label{aaa}
\end{equation}
for $n_{0}\gg 1$. Substituting Eq.~(\ref{aaa}) in Eq.~(\ref{aita}) we finally 
obtain
\begin{eqnarray}
\langle {\bf A}\rangle_{t} & = & a_{0}n_{0}-a_{0}\xi^{2}\left(\sin^{2}
\left(\frac{1}{2}\omega_{n_{0},n_{0}-1}t\right)-\sin^{2}\left(\frac{1}{2}
\omega_{n_{0},n_{0}+1}t\right)
\right)\nonumber\\
& = & a_{0}n_{0}-a_{0}\xi^{2}\sin(\omega^{(1)}t)\sin(\omega^{(2)}t)
\label{aps}
\end{eqnarray}
where (with $G$ and $c$ displayed)
\begin{equation}
\omega^{(1)}=\frac{\ln k}{8\pi M}\left(\frac{c^{3}}{G}\right),\;\;\;     
\omega^{(2)}=\frac{(\ln k)^{2}\hbar}{128\pi^{2}M^{3}}\left(\frac{c^{4}}{G^{2}}
\right)\;
\;{\rm and}\;\;\xi =\frac{\pi\gamma\exp\left(\frac{\alpha}{8n_{0}}\right)}
{240(\ln k)^{2}}.
\end{equation}
Note that $\xi$ is a dimensionless coefficient, whereas $\gamma$ was first 
introduced in Eq.~(\ref{avom}).

The frequency $\omega^{(1)}$ does not contain $\hbar$; hence we call it
the {\it classical} frequency of the BH. The second frequency, $\omega^{(2)}$, 
is the quantum
correction to $\omega^{(1)}$, and even for a microscopic BH with 
$\langle {\bf M}(0)\rangle=10^{12}$Kg, $\omega^{(2)}\approx
10^{-19}$sec$^{-1}$(where we assumed $\ln k\sim 1$). Thus
$\sin (\omega^{(2)}t)\approx 0$ for $t$ not much larger that the age of the 
universe. We conclude that a BH (not a primordial one) which is described 
initially by a pure state $|n_{0}\rangle$ will stay in this pure state since 
$\langle {\bf A}\rangle_{t}=a_{0}n$ over the lifetime of the 
universe. 

The frequency $\omega^{(2)}$ becomes significant when $1/\omega^{(2)}$ is 
smaller then the age of the universe. That is, BHs with a mass smaller than
$10^{11}$Kg. For these BHs it is impossible to neglect $\omega^{(2)}$. But 
still, if we assume that $\xi$ (or $\gamma$) is not extremely large,
Eq.~(\ref{aps})  describes a basically static BH with small time dependent
fluctuations around its mean area value. 

\section{Summary and Conclusions}\label{sec:summary} 

We have described in this paper a fully quantum mechanical model for a
Schwarzschild BH. Starting from the assumption that the horizon area is
quantized with equally spaced spectrum, we have obtained for the black hole
observables an algebra of angular momentum (the hyperspin of a Schwarzschild
BH) with the restriction that the hyperspin can only have the values
$s_{n}=(k^{n}-1)/2$. This serves as motivation for the  introduction in the
Hamiltonian of an interaction term. Surprisingly, this term leads to the correct
properties of the HR. 

We have introduced the {\it phase} of a BH as the canonical conjugate to the 
area operator, and identified it with the ``opening angle"  of Carlip and
Teitelboim~\cite{CarTe95}. Then the ``interaction" term in the Hamiltonian
has been interpreted as a coupling  between the horizon area and the phase of
the BH. This coupling we associated  with the boundary term $\Theta A/8\pi$
in the action of a BH spacetime, which according to Massar and
Parentani~\cite{MasPar99} corresponds to the HR.

To discuss the HR it was required to choose an appropriate initial state
which most closely imitates the semi-classical result (HR). We have chosen the
initial state to be Gaussian distributed with area about some mean value 
$n_{0}$. Hence, the variance $\sigma_{A}$ of this state was left undetermined.
However, after the comparison with the HR, it has been shown that either 
$\sigma_{A}/\langle {\bf A}\rangle\sim {\cal M}_P/M$, or  
$\sigma_{A}/\langle {\bf A}\rangle\sim ({\cal M}_P/M)^{2}$. 

For the first possibility,  the Hamiltonian which describes the system is {\it
exactly} of the form of  Eq.~(\ref{newhamil}). On the other hand, if 
$\sigma_{A}/\langle {\bf A}\rangle\sim ({\cal M}_P/M)^{2}$, then the
Hamiltonian given by Eq.~(\ref{newhamil}) is just part (in a perturbation
theory) of the full Hamiltonian of the BH (see Appendix B). 

We have seen in the previous section that an area eigenstate describes a
static BH. Hence, in our model the HR is due to the dispersion in the area
eigenstates. It was shown long ago by Bekenstein~\cite{Beke84} that the
fluctuations in the mass of the BH, and thus also in its area, increases with
time.  Also Wu and Ford~\cite{WuFord99} have shown recently that the variance
of the mass grows  linearly in time. It is then clear that as long as the
BH looses mass, the observers loose their knowledge about the area of the 
BH. Perhaps this is the way that a BH avoids a singularity at the culmination
of its evaporation. 

\section*{Acknowledgments}
I would like to thank
Prof.~J.~Bekenstein for his guidance and
support during the course of this work.
They proved invaluable to me. It is also a pleasure
to thank A.~E.~Mayo, S.~Hod and L.~Sriramkumar
for helpful discussions. This research was supported
by a grant from the Israel Science Foundation, established 
by the Israel National Academy of Sciences.  
 
\appendix

\section{Setting {\it u}$({\bf N})=1$}

In order to prove that $u({\bf N})=1$ in Eq.~(\ref{ggd}), 
we first define three operators in the same
manner as in Eq.~(\ref{jjj}), that is
\begin{eqnarray}
{\bf J_{1}} & \equiv & \frac{1}{2}(l({\bf N}) {\bf g}+{\bf g^{\dag}}l^{*}({\bf N}))
\nonumber\\
{\bf J_{2}} & \equiv & \frac{i}{2}(l({\bf N}) {\bf g}-{\bf g^{\dag}}l^{*}({\bf N}))
\nonumber\\
{\bf J_{3}} & \equiv & |l({\bf N})|^{2}\left({\bf G}-\frac{1}{2}k^{{\bf N}}-\frac{1}{2}\right)
\label{ap:jjj}
\end{eqnarray}
where $u({\bf N})\equiv |l({\bf N})|^{2}$ since $u({\bf N})$ must have real
positive eigenvalues (see Eq.~(\ref{ggd})). 
These three operators satisfy the angular momentum commutation rules and thus the 
eigenvalues of ${\bf J_{3}}$ and ${\bf J}^{2}$ are $m_{3}=-j,-j+1,...,j$ and $j(j+1)$,
respectively, where $j$ is an integer or half integer.

Using Eq.~(\ref{ap:jjj}) and Eq.~(\ref{ggd}) one can obtain
\begin{equation}
4{\bf J}^{2}=|l({\bf N})|^{2}\left(k^{2{\bf N}}-1-\left(1-|l({\bf N})|^{2}\right)
\left(2{\bf G}-k^{\bf N}-1\right)^{2}\right).
\label{ap:rjng}
\end{equation}
Hence it is clear that the eigenstates $|n,m\rangle$ of ${\bf N}$ and ${\bf G}$
are also eigenstates of ${\bf J}_{3}$ and ${\bf J}^{2}$. Thus we shall write $m_{3}$
and $j$ in terms of $n$ and $m$ as follows:
\begin{eqnarray}
2m_{3} & = & |l(n)|^{2}\left(2m-k^{n}-1\right)\label{ap:a}\\
4j(j+1) & = & |l(n)|^{2}\left(k^{2n}-1-\left(1-|l(n)|^{2}\right)
\left(2m-k^{n}-1\right)^{2}\right).\label{ap:b}
\end{eqnarray}
From Eq.~(\ref{ap:a}) it follows that $|l(n)|^{2}$ is an integer since $2m_{3}$ is an 
integer. Thus, $|l(n)|^{2}\geq 1$. 

Setting for example $m=1$ in 
Eq.~(\ref{ap:a}) and Eq.~(\ref{ap:b}) we find that
\begin{eqnarray}
2m_{3} & = & |l(n)|^{2}\left(1-k^{n}\right)\label{ap:am}\\
4j(j+1) & = & |l(n)|^{2}\left(k^{n}-1\right)
\left(k^{n}+2-|l(n)|^{2}\right).\label{ap:bm}
\end{eqnarray}
Now, the eigenvalues of an angular momentum satisfy the condition $2j\geq |2m_{3}|$
which leads to $4j(j+1)\geq |2m_{3}|\left(|2m_{3}|+2\right)$. Hence, using both 
Eq.~(\ref{ap:am}) and Eq.~(\ref{ap:bm}) we finally obtain
\begin{equation}
|l(n)|^{2}\left(k^{n}-1\right)\left(k^{n}+2-|l(n)|^{2}\right)
\geq |l(n)|^{2}\left(k^{n}-1\right)\left(k^{n}+1\right)
\end{equation}
which is satisfied {\it only} if $|l(n)|^{2}\leq 1$. Thus we have proved that 
$u(n)\equiv |l(n)|^{2}=1$.
 
\section{A Second Order Correction to the Hamiltonian}

The second order correction is related to transitions $n\rightarrow n\pm 2$
(see section~\ref{sec:gfh}), and hence may be written as:
\begin{equation}
{\bf H}^{(2)}=\sum_{n=0}^{\infty}\left (r(n)|n+2\rangle\langle n|+r^{*}(n) 
|n\rangle\langle n+2|\right)
\label{doda}
\end{equation}
where $r(n)$ is a complex function of $n$. Now, since to zeroth order
${\bf H}^{(0)}\equiv {\bf M}\sim\sqrt{\bf A}$, and the first order correction
${\bf H}^{(1)}\equiv {\bf V}\sim 1/\sqrt{\bf A}$ for large $\langle{\bf A}\rangle$
values, we conclude that for large $\langle{\bf A}\rangle$ values the second order
correction is proportional to $1/{\bf A}^{3/2}$, i.e. each term of a higher
order in the perturbation theory is smaller by one extra power of ${\bf A}$.
Thus, the function $r(n)$ in Eq.~(\ref{doda}) can be written for large $n$ as
\begin{equation}
r(n)=\zeta\frac{(\hbar\varepsilon)^{2}}{m_{0}^{5}}{\rm e}^{(\frac{\alpha}{8n_{0}})}
\frac{1}{n^{3/2}}
\end{equation}
where $\zeta$ is a dimensionless number. It is reasonable to assume that $\zeta$
is of the order of unity since the dimensionless coefficient of $r(n)/m_{0}$
in a perturbative theory is supposed to be of the same order as the square of the
coefficient of $f(n)/m_{0}$. We might also assume that $\zeta$ is purely
real (and also negative), but this is not necessary because the 
imaginary part of $\zeta$ (if it exists) will not contribute to Eq.~(\ref{devi}). 

We shall now take into account ${\bf H}^{(2)}$ in the expression
$[{\bf H},[{\bf H},{\bf M}(0)]]$ in Eq.~(\ref{sason}). The only term in 
$\langle[{\bf H},[{\bf H},{\bf M}(0)]]\rangle$
which contains ${\bf H}^{(2)}$, and is of the order of $1/\langle M(0)\rangle^{5}$,
is $\langle[{\bf M}(0),[{\bf H}^{(2)},{\bf M}(0)]]\rangle$. All the other
terms are smaller by a factor of $1/\langle M(0)\rangle^{2}$, and may thus be neglected.
A straightforward calculation gives
\begin{equation}
\langle[{\bf M}(0),[{\bf H}^{(2)},{\bf M}(0)]]\rangle_{CT;n_{0}}=2\zeta
\frac{(\hbar\varepsilon)^{2}}{\langle M(0)\rangle^{5}}e^{-\frac{\alpha}{8n_{0}}}.
\end{equation}
Hence, we are now able to rewrite Eq.~(\ref{devi}), including the second order correction
of the Hamiltonian, as
\begin{equation}
3{\rm e}^{(\frac{\alpha}{8n_{0}})}+(1+4\zeta){\rm e}^{(-\frac{\alpha}{8n_{0}})}=4
\label{newdevi}
\end{equation}
where $\zeta$ is still to be determined. Note that if we choose $\zeta=0$ 
(that is, ${\bf H}^{(2)}=0$) Eq.~(\ref{newdevi}) reduces to Eq.~(\ref{devi}), as it was
expected. Thus, we conclude that if $\alpha$ is of the order of unity, then 
Eq.~(\ref{devi}) is satisfied automatically and we do not need a correction. 
This means that the Hamiltonian given in Eq.~(\ref{hamil}) with the interaction 
term~(\ref{iaph}) is equivalent to the {\it exact} Hamiltonian, at least up to 
O($\hbar^{2}/\langle M(0)\rangle^{3}$). On the other hand, 
if $\alpha$ is O($n_{0}$), then the interaction term~(\ref{iaph}) is 
only the first order term in a given perturbation theory.

In order to determine $\alpha$ in this last case, we use Eq.~(\ref{newdevi}). 
There are two solutions
of Eq.~(\ref{newdevi}), but only one of them is plausible:
\begin{equation}
{\rm e}^{(\frac{\alpha}{8n_{0}})}=\frac{2+\sqrt{1-12l}}{3}.
\label{rela}
\end{equation}
Note that $l\leq 0$ since $\alpha>0$ as it was expected. Relation~(\ref{rela}) does not
determine $\alpha$, but instead express it as a function of the parameter $l$.

\section{Saturation of the Area-Phase Uncertainty Relation} 

We shall prove in this section that the CT-states defined in Eq.~(\ref{aphu})
saturate the area-phase uncertainty relation:
\begin{equation}
\Delta{\bf N}\Delta{\bf\phi}=\frac{1}{2}.
\label{snp}
\end{equation}
The variance of ${\bf N}$ for the Gaussian CT-states is given by
\begin{equation}
\Delta{\bf N}=\sqrt{\frac{2n_{0}}{\alpha}}.
\end{equation}
We now turn to calculate $\Delta{\bf\phi}$.

First we define $\sin{\bf\phi}$ in terms of the Susskind-Glogower's operators:
\begin{equation}
\sin{\bf\phi}=\frac{1}{2i}\left({\bf E}-{\bf E}^{\dag}\right).
\end{equation}
It is simple to verify that $\langle\sin{\bf\phi}\rangle_{CT;n_{0}}=0$; 
it is thus consistent to assume that also $\langle {\bf\phi}\rangle_{CT;n_{0}}=0$, where
${\bf\phi}$ has been chosen to have its eigenvalues in the interval $[-\pi,\pi]$.
If $\alpha\sim 1$, the variance $\Delta{\bf N}\sim\sqrt{n_{0}}$ is quite large, 
which implies that $\Delta{\bf\phi}$ is quite small. That is, ${\bf\phi}$ is highly
peaked around $\phi=0$. Hence we can assume that $\phi\approx\sin{\bf\phi}$ which 
leads by Eq.~(\ref{thst}) to 
\begin{equation}
\langle {\bf\phi}^{2}\rangle_{CT;n_{0}}\approx
\langle (\sin{\bf\phi})^{2}\rangle_{CT;n_{0}}=\frac{1}{2}\left(1-\exp
\left(-\frac{\alpha}{4n_{0}}\right)\right)\approx\frac{\alpha}{8n_{0}}
\end{equation}
for $\alpha\sim 1$. Hence,
\begin{equation}
\Delta{\bf\phi}=\frac{1}{2}\sqrt{\frac{\alpha}{2n_{0}}}
\end{equation}
and thus Eq.~(\ref{snp}) is satisfied.

For $\alpha\sim n_{0}$ it is impossible to assume that ${\bf\phi}$ 
is highly peaked around zero; instead we shall assume that the main contribution to 
$\langle {\bf\phi}^{2}\rangle_{CT;n_{0}}$ comes from the region where 
$-\pi/2<\phi<\pi/2$. In this interval ${\bf\phi}$ can be expressed as:
\begin{equation}
{\bf\phi}^{2}=\lim_{m\rightarrow\infty}\left({\bf\phi}^{2}\right)_{m}
\end{equation}
where
\begin{equation}
\left({\bf\phi}^{2}\right)_{m}\equiv\sum_{l=1}^{m}C_{l}\sin ^{2l}{\bf\phi}=
\sin ^{2}{\bf\phi}+\frac{1}{3}\sin ^{4}{\bf\phi}+\frac{8}{45}\sin ^{6}{\bf\phi}
+...+C_{m}\sin ^{2m}{\bf\phi}.
\end{equation}
The CT-states has the feature that
\begin{equation}
\langle ({\bf E})^{l}\rangle_{CT;n_{0}}=\langle ({\bf E}^{\dag})^{l}\rangle_{CT;n_{0}}
=\exp\left(-\frac{l^{2}\alpha^{*}}{16}\right)
\label{aved}
\end{equation}
where $l$ is an integer and $\alpha^{*}\equiv\alpha/n_{0}$ is of the order of unity. 
Hence, expressing 
$\left({\bf\phi}^{2}\right)_{m}$ as polynomials in ${\bf E}$ and ${\bf E}^{\dag}$ 
and then taking the average according to Eq.~(\ref{aved}), we were able to show 
(numerically) that
\begin{equation}
\langle \left({\bf\phi}^{2}\right)_{m}\rangle_{CT;n_{0}}=\frac{\alpha^{*}}{8}
+{\rm O}\left((\alpha^{*})^{m+1}\right)
\end{equation}
were we have expressed the exponential functions in Eq.~(\ref{aved}) as a Taylor series
of $\alpha^{*}$. Hence, we conclude that for $\alpha^{*}<1$ 
\begin{equation}
\langle {\bf\phi}^{2}\rangle_{CT;n_{0}}= \frac{\alpha^{*}}{8}=\frac{\alpha}{8n_{0}}
\end{equation}
which leads to Eq.~(\ref{snp}).


\begin{references}

\bibitem{Kuchar}K. Kucha\v r, gr-qc/9304012.

\bibitem{Weinberg}S. Weinberg, in {\it General Relativity: An Einstein
Centenary Survey\/}, eds. S. W. Hawking and W. Israel (Cambridge University
Press, Cambridge 1979).

\bibitem{dbrane}
M.~B.~Green, J.~H.~Schwarz and E.~Witten, {\it Superstring theory} (Cambridge
University Press, Cambridge 1987); J.~Polchinski, {\it String Theory} 
(Cambridge University Press,  Cambridge 1998).  

\bibitem{AshKra}
A.~Ashtekar and K.~Krasnov, gr-qc/9804039; A.~Ashtekar, C.~Rovelli and
L.~Smolin, Phys.\ Rev.\ Letters {\bf 69}, 237 (1992).

\bibitem{Bek74}
J.~D.~Bekenstein, Lett.\ Nuovo Cimento\ {\bf 11}, 467 (1974).

\bibitem{Mu86}V. Mukhanov, Pis. Eksp. Teor. Fiz. {\bf 44}, 50 (1986)
[JETP Letters {\bf 44}, 63 (1986)]

\bibitem{BekMu95}
J.~D.~Bekenstein and V.~F.~Mukhanov, Phys.\ Lett.\ B\ {\bf 360}, 7 (1995).

\bibitem{BekLectures}J.~D.~Bekenstein, gr-qc/9808028.

\bibitem{Chri}
D.~Christodoulou, Phys.\ Rev.\ Lett.\ {\bf 25}, 1596 (1970).

\bibitem{Haw71}
S.~W.~Hawking, Phys.\ Rev.\ Lett.\ {\bf 26}, 1344 (1971).

\bibitem{PenFly}
R.~Penrose and R.~M.~Floyd, Nature\ {\bf 229}, 177 (1971). 

\bibitem{Bek73}
J.~D.~Bekenstein, Phys.\ Rev.\ D\ {\bf 7}, 2333 (1973);


\bibitem{Bek98}
J.~D.~Bekenstein, in {\it Black Holes,
Gravitational Radiation and the Universe}, eds. B.~R.~Iyer and B.~Bhawal 
(Kluwer, Dordrecht 1998).

\bibitem{Mayo}
A.~E.~Mayo, Phys.\ Rev.\ D\ {\bf 58}, 104007 (1998).

\bibitem{Eren}
M.~Born, {\it Atomic Physics} (Blackie, London 1969), eight edition.

\bibitem{Hod1}
S.~Hod, Phys.\ Rev.\ D {\bf 59}, 024014 (1999).

\bibitem{Hawk75}
S.~W.~Hawking, Commun.\ Math.\ Phys.\ {\bf 43}, 199 (1975);\ Nature 
\ {\bf 248}, 30  (1974).

\bibitem{Kast97}
H.~A.~Kastrup, Phys.\ Lett.\ B\ {\bf 413}, 267 (1997).

\bibitem{CenLo98}
C.~Vaz and L.~Witten, Phys.\ Lett.\ B\ {\bf 442}, 90 (1998).

\bibitem{JorJar96}
J.~Louko and J.~M\"akel\"a, Phys.\ Rev.\ D\ {\bf 54}, 4982 (1996).

\bibitem{Kast96}
H.~A.~Kastrup, Phys.\ Lett.\ B\ {\bf 385}, 75 (1996).

\bibitem{CarTe95}
S.~Carlip and C.~Teitelboim, Class.\ Quant.\ Grav.\ {\bf 12}, 1699 (1995).

\bibitem{Lyn95}
R.~Lynch, Phys.\ Rep.\ {\bf 256}, 367 (1995).

\bibitem{Hod2}
S.~Hod, Phys.\ Rev.\ Lett.\ {\bf 81}, 4293 (1998).

\bibitem{Pag76}
D.~N.~Page, Phys.\ Rev.\ D\ {\bf 13}, 198 (1976).

\bibitem{Sakurai}
J.~J.~Sakurai, {\it Modern Quantum Mechanics} (Addison Wesley Press,
 New York 1995).

\bibitem{NoHair}
R.~Ruffini and J.~A.~Wheeler, Physics Today\ {\bf 24}, 30 (1971).

\bibitem{BKSS99}
M.~Bojowald, H.~A.~Kastrup, F.~Schramm and T.~Strobl, gr-qc/9906105.

\bibitem{Hawki75}
S.~W.~Hawking, Phys.\ Rev.\ D\ {\bf 14}, 2460 (1975).

\bibitem{Gour99}
G.~Gour, gr-qc/9907066 (To appear in Physical Review D). 

\bibitem{Padi}
T.~Padmanabhan, Phys.\ Lett.\ A\ {\bf 136}, 203 (1989).

\bibitem{KLJ} 
H.~C.~Kim, M.~H.~Lee and J.~Y.~Ji, gr-qc/9709062.

\bibitem{SusGlo64}
L.~Susskind and J~.Glogower, Physics\ {\bf 1}, 49 (1964).

\bibitem{MasPar99}
S.~Massar and R.~Parentani, gr-qc/9903027.

\bibitem{Umeza}
H.~Umezawa, {\it Advanced Field Theory} (AIP Press, New York, 1995). 

\bibitem{Beke84}
J.~D.~Bekenstein, in {\it Quantum Theory of Gravity}, ed.~S.~M.~Christensen
(Adam Hilger, Bristol 1984).

\bibitem{WuFord99}
C.~H.~Wu and L.~H.~Ford, gr-qc/9905012.



\end{references}
\end{document}